\documentclass[a4paper,fleqn,usenatbib]{mnras}
\usepackage[T1]{fontenc}
\usepackage{ae,aecompl}
\usepackage{graphicx}	
\usepackage{amsmath}	
\usepackage{amssymb}	
\usepackage{amsbsy}
\usepackage{float}
\usepackage{color}
\usepackage{booktabs,caption}

\usepackage{mathrsfs,bm}
\newcommand{\bcdot}{\ensuremath{%
  \mathchoice%
   {\mskip\thinmuskip\lower0.2ex\hbox{\scalebox{1.5}{$\cdot$}}\mskip\thinmuskip}}%
   {\mskip\thinmuskip\lower0.2ex\hbox{\scalebox{1.5}{$\cdot$}}\mskip\thinmuskip}%
   {\lower0.3ex\hbox{\scalebox{1.2}{$\cdot$}}}%
   {\lower0.3ex\hbox{\scalebox{1.2}{$\cdot$}}}%
}

\def\del#1{{}}

\newcommand{\Msun}{M$_\odot$}

\newcommand{\R}{$R_{200}$}
\title[Quenching and stripping of satellite galaxies]{Quenching and ram pressure stripping of simulated Milky Way satellite 
galaxies}

\author[C.M.~Simpson et al.]  {Christine~M.~Simpson$^1$\thanks{E-mail: Christine.Simpson@h-its.org}, Robert J. J. Grand$^{12}$, 
Facundo A. G\'{o}mez$^3$, \newauthor Federico Marinacci$^4$, R\"{u}diger Pakmor$^1$, Volker Springel$^{12}$,  \newauthor David J. R. 
Campbell$^5$ and Carlos S. Frenk$^5$\vspace{10pt} \\
 $^1$Heidelberger Institut f\"{u}r Theoretische Studien, Schloss-Wolfsbrunnenweg 35, 69118 Heidelberg, Germany\\
 $^2$Zentrum f\"{u}r Astronomie der Universit\"{a}t Heidelberg, Astronomisches Recheninstitut, M\"{o}nchhofstr. 12-14, 69120 Heidelberg, 
 Germany\\
 $^3$Max-Planck-Institut f\"{u}r Astrophysik, Karl-Schwarzschild-Str. 1, D-85748, Garching, Germany  \\
 $^4$Department of Physics, Kavli Institute for Astrophysics and Space Research, MIT, Cambridge, MA 02139, USA \\
 $^5$Institute for Computational Cosmology, Department of Physics, Durham University, South Road, Durham, DH1 3LE, UK\\
}
 
\pubyear{2017}

\begin{document}

\label{firstpage}
\pagerange{\pageref{firstpage}--\pageref{lastpage}}

\maketitle

\begin{abstract}
We present predictions for the quenching of star formation in satellite galaxies of the Local Group from a suite of 30 cosmological zoom simulations of 
Milky Way-like host galaxies.  The Auriga simulations resolve satellites down to the luminosity of the classical dwarf spheroidal galaxies of the Milky Way.  
We find strong mass-dependent and distance-dependent quenching signals, where dwarf systems beyond 600 kpc are only strongly quenched below a 
stellar mass of $10^7$ \Msun.  Ram pressure stripping appears to be the dominant quenching mechanism and 50\% of quenched systems cease star 
formation within 1 Gyr of first infall.  We demonstrate that systems within a host galaxy's \R\ radius are comprised of two populations: (i) a first infall 
population that has entered the host halo within the past few Gyrs and (ii) a population of returning `backsplash' systems that have had a much more 
extended interaction with the host.  Backsplash galaxies that do not return to the host galaxy by redshift zero exhibit quenching properties similar to galaxies 
within \R\ and are distinct from other external systems.  The simulated quenching trend with stellar mass has some tension with observations, but our 
simulations are able reproduce the range of quenching times measured from resolved stellar populations of Local Group dwarf galaxies. 
\end{abstract}
 

\section{Introduction}

The Milky Way (MW) galaxy and M31 host systems of satellite dwarf galaxies that have been of great interest in recent decades because 
of the perceived challenges they pose to $\Lambda$CDM in terms of their number \citep{Klypin1999,Moore1999}, 
structure \citep{BoylanKolchin2011}, and spatial distribution \citep{Ibata2013}.  These systems also pose challenges to models of galaxy 
formation, as many appear to be some of the most dark matter dominated objects in the Universe \citep{McGaugh2010}.  Studies of 
resolved stellar populations in these systems have shown a diversity of star formation histories and provided a local window into the 
high-redshift Universe by constraining these histories over 12 Gyr \citep[e.g.][]{Weisz2014}.  A holistic cosmological model of the Local 
Group needs not only to explain the many structural properties of satellite galaxies, but also their star forming properties over a Hubble time.

Properties of dwarf galaxies in the Local Group exhibit strong trends with both luminosity and environment.  There is a clear 
metallicity-luminosity trend that may indicate strong galactic winds in these systems \citep{Kirby2011}.  There is a strong correlation 
between distance from either the MW or M31 and gas content.  Few satellites within 270 kpc of either the MW or M31 contain detectable 
H~I gas and most systems beyond this distance do contain H~I \citep{Grcevich2009}, sometimes in significant 
amounts \citep{RyanWeber2008}.  There is also a strong correlation between distance and galaxy type in the Local Group \citep{Grebel1999} 
that may connect to distance trends for intermediate luminosity dwarfs (with stellar masses above $10^7$ \Msun), where most systems in the field 
are found to be star-forming and of late type \citep{Geha2012}.  A small population of early type field dwarfs are known, although at lower 
luminosities \citep{Makarova2017}.

Within the Local Group, the satellite populations of the MW and M31 exhibit some striking differences.  For example, M31 contains several 
dwarf elliptical (dE) galaxies, while the MW has none \citep{McConnachie2012}.  Also, M31 contains several dwarf spheroidal (dSph) 
galaxies with unusually large half-light radii compared to MW dSphs \citep{Collins2013}.  The MW contains the Magellanic clouds, two 
nearby, gas rich and actively star-forming dwarf irregular (dIrr) galaxies, that are the main exceptions to the distance trends otherwise 
observed in the Local Group.  There are also differences in the globular cluster populations around the MW and M31 \citep{Huxor2011}.  A 
possible explanation for these differences is cosmic variance, i.e. the statistical variance in the dark matter assembly histories of two 
otherwise similar dark matter haloes.  A theoretical exploration (e.g. in simulations) of satellite properties in the context of the Local Group 
therefore needs to account for cosmic host variance. 

To date, most simulation studies of MW analog satellite systems have typically focused either on one to a few high resolution cosmological 
models with gas physics \citep{Okamoto2010,Zolotov2012,Wetzel2016} or have focused on a large suite of dark matter only models that 
better probe cosmic variance, but do not include the effects of baryons \citep{GarrisonKimmel2014,Hellwing2016}.  An exception is the APOSTLE 
simulations, a set of 12 cosmological zoom simulations of Local Group-like volumes \citep{Fattahi2016,Sawala2016a,Sawala2016b}.  These 
simulations have demonstrated the impact of reionization and tidal stripping on dwarf satellites with a model that includes both gas physics and dark matter 
across a large sample of dwarf satellite systems.

The physics that shapes the star formation histories of satellite galaxies is diverse, but interaction with the host galaxy likely plays an 
important role through both tidal effects and ram pressure stripping \citep{Gatto2013,Emerick,Zhu2016}.  The shallower gravitational potentials of these 
systems boost the impact of feedback processes and, at very low masses, reionization likely plays an important role
\citep{Efstathiou1992,ThoulWeinberg1996,Benson2002}.

The goal of this study is to identify the physical mechanisms that drive quenching in Local Group satellite galaxies at the luminosity scale of 
the `classical' dSphs, which are well studied observationally.  To do this, we will use the simulations of the Auriga project, a suite of 30 cosmological, 
hydrodynamical simulations of Milky Way-like galaxies \citep{Grand2016}.  Our aim will be to quantify patterns in quenching and gas loss 
among subhaloes within a local volume around the simulated host galaxies ($< 1$ Mpc) and explore the processes, such as ram pressure 
stripping, that shape these patterns.  We will then quantify timescales for gas loss and quenching and compare these timescales to the 
infall histories of systems  in order to better understand how the final quenching patterns emerge over time.  The perspective of this study is 
that of the surviving, luminous satellite dwarf galaxy.  Systems that are tidally destroyed before $z=0$ and systems that are dark or have 
under-resolved stellar populations are also interesting, but will be addressed in later work.

The structure of this paper is as follows.  In Section \ref{sec:methods}, we describe the Auriga simulations and their galaxy formation 
model.  In Section \ref{sec:properties}, we examine the properties of surviving luminous subhaloes across the simulation sample and in 
Section \ref{sec:processes} we turn to the physical processes responsible for these properties, with a particular focus on ram pressure 
stripping.  Section \ref{sec:timescales} contains an exploration of timescales associated with quenching and the connection between 
quenching and  subhaloes' orbital histories.  In Section \ref{sec:discussion}, we discuss how our results compare to observations and we 
present conclusions in Section \ref{sec:conclusions}.

\section{Methods}
\label{sec:methods}

The thirty `Level 4' simulations of the Auriga project \citep{Grand2016} are the simulation sample we use to investigate dwarf galaxy 
quenching in the Local Group.  Here we describe the Auriga simulations' set-up and physical model and the analysis techniques, such as 
merger tree finding, that we use for later analysis.

\subsection{The Auriga Simulations}

The Auriga project consists of cosmological zoom-in simulations of $\sim 10^{12}$ \Msun\ dark matter haloes within the $\Lambda$CDM 
paradigm conducted with the second-order accurate moving mesh code AREPO \citep{Springel2010,Pakmor2016}.  All simulations include 
collisionless dark matter; ideal magneto-hydrodynamics with a Powell cleaning scheme \citep{Pakmor2011,Powell1999}; primordial and 
metal-line cooling \citep{Vogelsberger2013}; a sub-grid model for the interstellar medium (ISM) that employs a stiff equation of state 
representing a two-phase medium in pressure equilibrium \citep{SpringelHernquist2003}; a model for star formation and stellar feedback, 
including a phenomenological wind model and metal enrichment from SNII, SNIa and AGB stars \citep{Vogelsberger2013}; the formation of 
black holes and feedback from active galactic nuclei \citep{Springel2005, Marinacci2014, Grand2016}; and a spatially uniform but time 
varying UV background that completes reionization at redshift six \citep{FaucherGiguere2009,Vogelsberger2013}.

Each Auriga host halo was selected from a cosmological dark matter only simulation of the EAGLE project \citep{Schaye2015}. 
The parent cube has a side length of 100 comoving Mpc and cosmological parameters $\Omega_m = 0.307$, $\Omega_b = 0.048$, 
$\Omega_\Lambda = 0.693$, and $h=0.6777$ \citep{Planck2014}.  Candidate haloes were required to fall within a narrow range of final 
virial halo mass around $10^{12}$ \Msun.  

\begin{figure*}
\centering
\includegraphics[width=\textwidth]{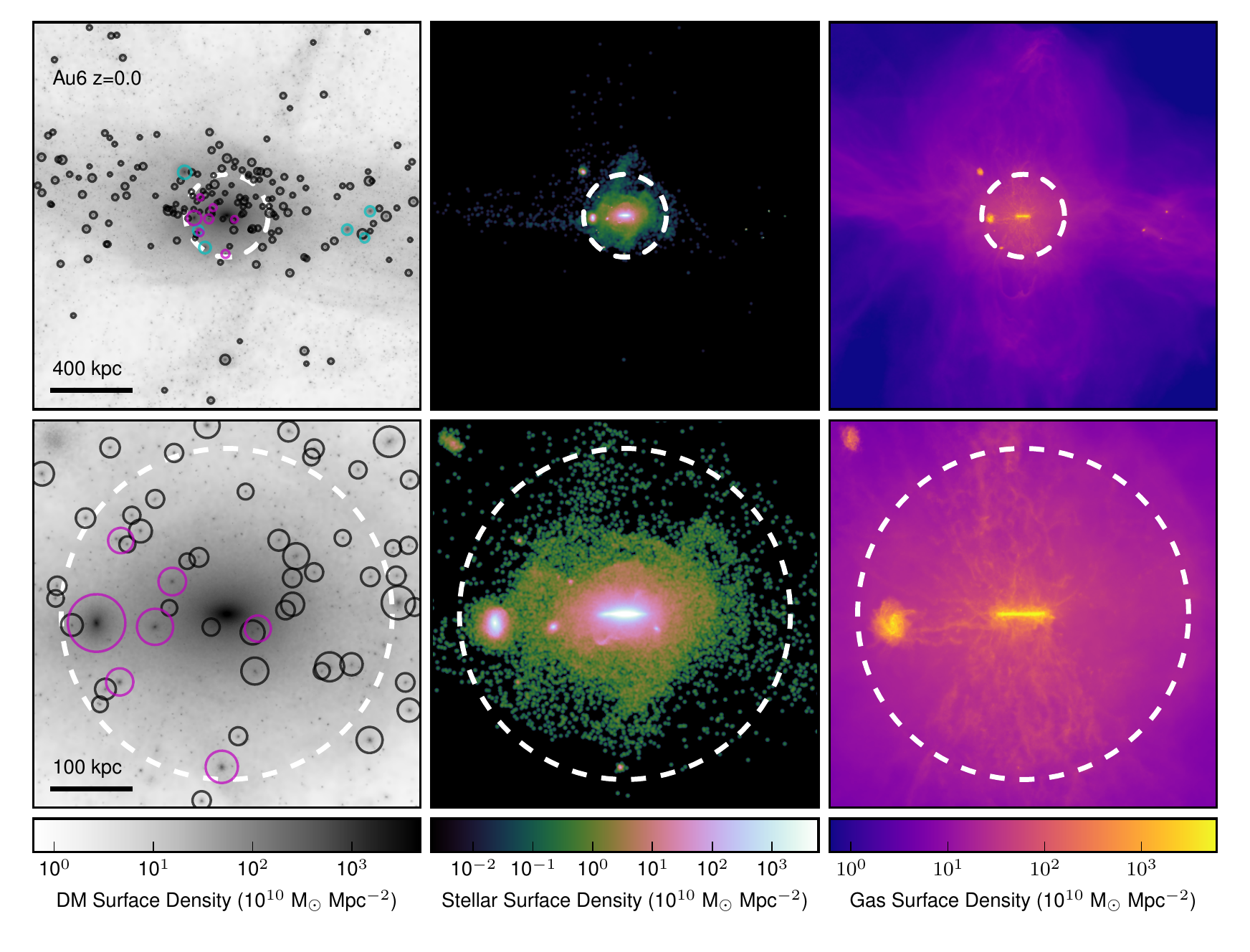}
\caption{Projections of an example host halo (Au6) and nearby low-mass subhaloes at redshift zero.  The top row shows a wide view 
that is 2 Mpc across; the bottom row shows a more zoomed-in view that is 500 kpc across.  The depth of each projection is equal to 
the width.  In each panel, a white dashed circle shows the \R\ radius (214 kpc).  The projections are oriented to view the disk of the 
host halo edge-on.  The left column shows projections of dark matter density.  Every subhalo with a total mass greater than $10^8$ \Msun\ 
is circled.  Subhaloes that have a stellar mass greater than $5 \times 10^5$ \Msun\ are circled in color: magenta for subhaloes within the 
host \R, and cyan for subhaloes beyond \R.  The diameter of each subhalo circle is proportional to the logarithm of the subhalo's mass.
The middle column shows projections of stellar density and the right column shows projections of gas density.  The zoomed-out view of the 
top row reveals that there is a large network of low mass subhaloes out to large distances ($\sim$ 1 Mpc) and many of these subhaloes are 
completely devoid of gas and stars.     }
\label{fig:host_image}
\end{figure*}

A weak isolation criterion was used to select candidate haloes that required the closest halo of equivalent or larger mass to be at least 9 \R\ 
lengths distant (\R\ is defined as the radius within which the halo's mean density is equal to 200 times the critical density of the universe).  
This isolation criterion did not impose any restrictions on selected haloes' substructure.  In some cases, the selected haloes host massive 
subhaloes that contain large galaxies themselves (for example, Au1 has a nearly equal mass companion within \R).  The process for host halo 
selection is described in more detail by \citet{Grand2016}.

Zoom initial conditions were created for each selected Auriga host halo following the procedure of \citet{Jenkins2010}.  The region of 
interest, containing the host halo and a surrounding buffer region with many other smaller haloes, is simulated with lower-mass dark 
matter particles, while the rest of the box is simulated with higher-mass dark matter particles.  In most of the thirty zoom simulations, 
the closest low resolution particle to the host halo is more than 4\R\ distant and, in all cases, more than 3\R\ distant.  

High-resolution dark matter particles have a mass of $\sim 3 \times 10^5$ \Msun.  Gas cells have a target mass of $\sim 5 \times 10^4$ 
\Msun, and therefore, both gas cells and star particles have masses within a factor of 2 of this value.  Gravitational forces for star 
particles and high-resolution dark matter particles are computed with a comoving softening length of 500 $h^{-1}$pc until $z=1$, 
after which a constant physical softening length of 396 pc is used.  Gravitational forces for gas cells are computed with a softening 
length that scales with the mean cell radius, but limited to be between the star particle softening length and 1.85 physical kpc.

\subsection{Simulation Data and Analysis}

All thirty Auriga simulations are denoted by the descriptor `AuN' where N varies between 1 and 30.  Each simulation produces 128 full 
snapshots with contemporaneous halo catalogs that have a maximum spacing between snapshots of 167 Myr that decreases toward 
higher redshift.  All simulations were run with the in-simulation halo finder SUBFIND \citep{Springel2001} that identifies central haloes 
with a friends-of-friends (FOF) algorithm \citep{Davis1985} and gravitationally bound subhaloes.  

The spherical region uncontaminated by low-resolution dark matter particles 
extends out to 4\R\ in most of the simulations; however, due to the web-like structure of dark matter on the scale of Mpcs, the high-resolution Lagrangian 
region is irregular and almost all of the haloes and subhaloes with 1 Mpc of the main halo are uncontaminated.  We judge whether a subhalo is contaminated 
by measuring the number of low-resolution dark matter particles within 3$r_{200}^{\rm{subhalo}}$, where $r_{200}^{\rm{subhalo}}$ is defined for each subhalo 
as the radius within which the average subhalo density is 200 times the critical density of the universe in the same way \R\ is defined for the host halo.  Any 
subhalo that has coarse resolution dark matter particles within 3$r_{200}^{\rm{subhalo}}$ is considered to be contaminated.  In all thirty simulations, only one 
subhalo within 1 Mpc at $z=0$ in one simulation (Au15) was found to be contaminated.  We exclude it from our analysis.

In post-processing, merger trees were constructed with the LHaloTree algorithm that creates a merger linking between 
successive halo catalogs, including subhaloes of central haloes \citep{Springel2005Mill}.  The version of LHaloTree applied here includes a 
modification of the traditional algorithm that uses a cumulative sum of all of a subhalo's ancestors' masses (computed recursively) to 
distinguish the primary progenitor among all progenitors of a given generation.  This improvement reduces discontinuous jumps in 
properties along a halo's main ancestral line \citep{DeLucia2007}.

For the purpose of visualizing some of the fast moving processes that operate on timescales less than 200 Myr, such as ram pressure 
stripping of satellites, we re-ran Au6 with the same mass resolution, but with an increased number of snapshots (1024 outputs rather than 
128).  This resimulation has a maximum time between successive snapshots of 21 Myr.  The satellite distribution in this simulation has 
some small differences in the orbital histories of the satellites due to the differences in the time-step hierarchy necessitated by the higher 
frequency of outputs, but the overall satellite distribution is statistically similar to that of the original simulation. The visualizations presented 
in Figure \ref{fig:stripping_image} were created with this simulation.  
 
\section{Properties of surviving subhaloes}
\label{sec:properties}

In all thirty of the Auriga zoom simulations, the host halo contains both luminous and dark subhaloes at redshift zero; here we describe the
overall number and general baryon properties of these structures.  Figure \ref{fig:host_image} shows the environment around an example
host (in Au6) and the distribution of subhaloes found within 1 Mpc.  First, it is evident that the host halo lies at the intersection of several 
dark matter filaments and that suhaloes preferentially lie along these filaments.  Second, it is clear from the projections of star particles and 
gas that most of these subhaloes are dark and gas poor.   And third, the complex structures in stars at distances of tens to hundreds of kpc 
from the host's disk indicate that many subhaloes have been tidally destroyed and the population of subhaloes examined here are just the 
survivors of a continuous process of subhalo accretion and destruction.  We will discuss the physical mechanisms that result in these 
properties and associated timescales in Sections \ref{sec:processes} and \ref{sec:timescales}.

Table \ref{tab} lists general properties the subhalo systems at redshift zero, such as the number of luminous subhaloes within 300 kpc and 
1 Mpc, the fraction of `quenched', non-star forming subhaloes, and the fraction of gas poor subhaloes.  These properties, and all 
subsequent analysis, are restricted to subhaloes that have a minimum total mass of $10^8$ \Msun, a minimum stellar mass of 
$5\times10^5$ \Msun\ within twice the half stellar mass radius ($r_{1/2}$), and a minimum dark matter mass of $10^6$ \Msun.  This last 
requirement is intended to exclude from our analysis starforming gas clumps spuriously detected by the SUBFIND algorithm, which aims to 
identify dark matter structures.  When referring to a subhalo's stellar mass, we mean the 
stellar mass within 2$r_{1/2}$ and when referring to a subhalo's total mass, we mean the mass of all resolution elements linked to 
the subhalo by SUBFIND.

The stellar mass cut that we have applied is ten times the target gas mass and is intended to ensure that the stellar populations 
considered are sampled by approximately ten star particles or more.  Of systems that meet the stellar mass threshold, only 5\% do not 
also meet the total mass threshold of $10^8$ \Msun.  These systems appear to have undergone massive tidal stripping and tend to be 
found closer to the host halo center, but are not numerous enough to alter the population trends explored here and in later sections.

The Auriga model seeds black holes with masses of $10^5$ \Msun\ $h^{-1}$ in subhaloes with masses of $5 \times 10^{10}$ \Msun\ $h^{-1}$ 
\citep{Grand2016}.  This model results in a small population of black holes in satellite galaxies.  For the subhalos that meet our resolution cuts within 1 Mpc, at 
redshift zero only 33 systems contain black holes (5\% of systems).   Black holes are only found in subhaloes with final stellar masses above $10^9$ \Msun\ 
and total subhalo masses above $2 \times 10^{10}$ \Msun.  Subhaloes with halo masses below the halo mass limit for seeding black holes had higher halo 
masses in the past prior to accretion into the main host.  Subhalo black holes have masses between $10^5$ \Msun\ and $5 \times 10^7$ \Msun.  At these 
masses, we expect feedback from these black holes to be negligible.  There is one subhalo (within Au-30 at a distance of 400 kpc) that contains a 
$\sim 3 \times 10^8$ \Msun\ black hole that may contribute to the evolution of the satellite galaxy.  This `satellite' galaxy is close in size to the main halo and is 
not a true dwarf galaxy.  There are some subhaloes with stellar and halo masses above $10^9$ \Msun\ and $2 \times 10^{10}$ \Msun\ respectively that do not 
have black holes because their peak halo mass never reached the black hole seeding limit.   

\begin{table*}
\caption{Subhalo population properties by host halo.}
\label{tab}
\begin{tabular}{lcccccc}
\hline
Simulation & Host $M_{200}$     & Host $R_{200}$ & $N_{\rm{sub}}$  & max($v_{\rm{max}}$) & $f_{\rm{quenched}}$    
  & $f_{\rm{HIpoor}}$\\
 & ($10^{12}$ \Msun) & (kpc)                   &   (< 300 kpc/1 Mpc)                & (km s$^{-1}$)       &  (< 300 kpc/1 Mpc)   
 &(< 300 kpc/1 Mpc)   \\
  \hline
 
 Au1 & 0.93 & 206 & 11/30 & 112 & 0.36/0.53 & 0.27/0.4 \\ 

Au2 & 1.91 & 262 & 13/25 & 152 & 0.62/0.44 & 0.62/0.44 \\ 

Au3 & 1.46 & 239 & 10/14 & 99 & 0.8/0.71 & 0.8/0.64 \\ 

Au4 & 1.41 & 236 & 13/31 & 50 & 0.77/0.58 & 0.69/0.48 \\ 

Au5 & 1.19 & 223 & 15/19 & 75 & 0.93/0.89 & 0.87/0.79 \\ 

Au6 & 1.04 & 214 & 7/13 & 90 & 0.86/0.69 & 0.71/0.38 \\ 

Au7 & 1.12 & 219 & 8/13 & 38 & 0.62/0.46 & 0.62/0.46 \\ 

Au8 & 1.08 & 216 & 10/26 & 160 & 0.8/0.58 & 0.7/0.46 \\ 

Au9 & 1.05 & 214 & 11/16 & 29 & 1.0/0.88 & 1.0/0.81 \\ 

Au10 & 1.05 & 214 & 11/18 & 66 & 0.55/0.56 & 0.36/0.28 \\ 

Au11 & 1.65 & 249 & 12/22 & 158 & 0.42/0.32 & 0.5/0.27 \\ 

Au12 & 1.09 & 217 & 7/14 & 65 & 0.57/0.36 & 0.57/0.29 \\ 

Au13 & 1.19 & 223 & 14/27 & 97 & 0.79/0.59 & 0.79/0.52 \\ 

Au14 & 1.66 & 249 & 7/17 & 64 & 0.71/0.76 & 0.57/0.59 \\ 

Au15 & 1.22 & 225 & 18/21 & 135 & 0.61/0.52 & 0.61/0.52 \\ 

Au16 & 1.5 & 241 & 15/30 & 117 & 0.87/0.7 & 0.73/0.53 \\ 

Au17 & 1.03 & 213 & 12/17 & 53 & 0.58/0.65 & 0.5/0.47 \\ 

Au18 & 1.22 & 225 & 13/44 & 84 & 0.62/0.57 & 0.62/0.52 \\ 

Au19 & 1.21 & 225 & 11/18 & 102 & 0.55/0.44 & 0.45/0.39 \\ 

Au20 & 1.25 & 227 & 16/30 & 133 & 0.75/0.57 & 0.69/0.5 \\ 

Au21 & 1.45 & 239 & 15/31 & 92 & 0.87/0.55 & 0.67/0.39 \\ 

Au22 & 0.93 & 205 & 6/8 & 75 & 0.67/0.62 & 0.67/0.5 \\ 

Au23 & 1.58 & 245 & 12/17 & 49 & 0.92/0.82 & 0.83/0.71 \\ 

Au24 & 1.49 & 241 & 13/37 & 103 & 0.85/0.59 & 0.77/0.46 \\ 

Au25 & 1.22 & 225 & 16/23 & 147 & 0.69/0.7 & 0.62/0.52 \\ 

Au26 & 1.56 & 245 & 15/29 & 45 & 0.93/0.72 & 0.93/0.69 \\ 

Au27 & 1.75 & 254 & 14/20 & 94 & 0.71/0.6 & 0.71/0.55 \\ 

Au28 & 1.61 & 247 & 12/25 & 46 & 0.83/0.6 & 0.58/0.36 \\ 

Au29 & 1.54 & 244 & 7/13 & 139 & 0.86/0.54 & 0.71/0.46 \\ 

Au30 & 1.11 & 218 & 12/25 & 145 & 0.92/0.76 & 0.92/0.72 \\ 

 \hline     

 \end{tabular}

\raggedright  Note: The quantities presented in each column are (1) the simulation name, (2) the host halo $M_{200}$, (3) the host halo 
$R_{200}$, (4) the number of satellites, defined as subhaloes within 300 kpc of the host halo (first number) or 1 Mpc (second number), with 
a minimum of stellar mass within $2r_{1/2}$ of $5 \times 10^5$ \Msun, a total subhalo mass greater than $10^8$ \Msun, and a dark matter 
mass greater than $10^6$ \Msun, (5) the $v_{\rm{max}}$ of the largest subhalo within 300 kpc, (6) the fraction of quenched 
subhaloes (see Section \ref{sec:sfproperties} for a description of our quenching criteria), and (7) the fraction of subhaloes that have less 
than $10^5$ \Msun\ in H~I mass.  These simulations are the Level 4 resolution simulations from \citet{Grand2016}.

\end{table*}

\subsection{Luminosity distributions}
\begin{figure*}
\centering
\includegraphics[width=\textwidth]{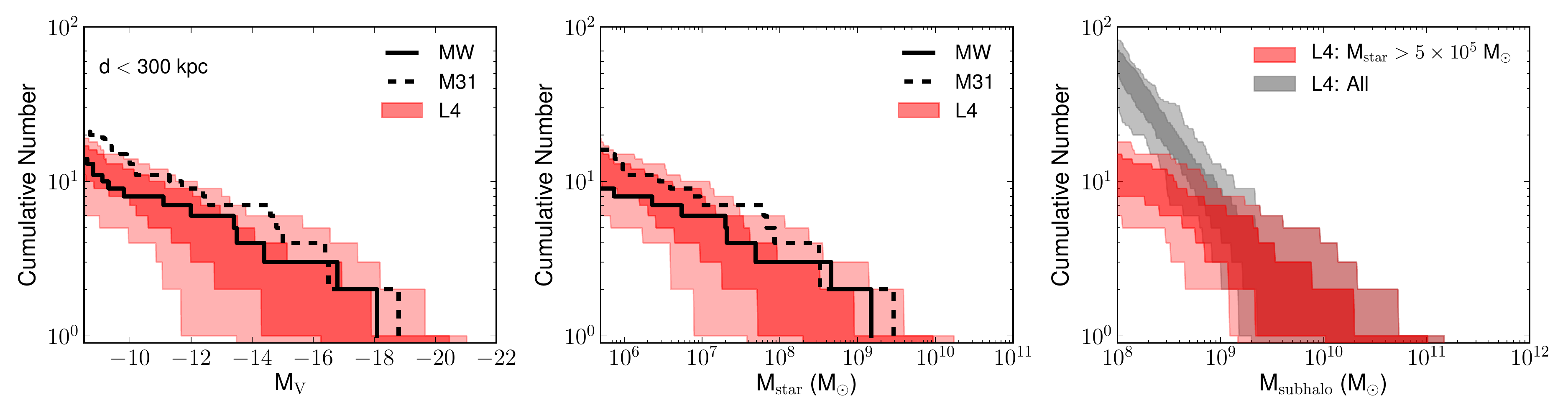}
\caption{Cumulative distributions of satellite galaxy properties for systems within 300 kpc of each simulated host.  Left: Ensemble of  
cumulative distributions of satellite V band absolute magnitudes, $M_V$.   The light red envelope encloses the individual cumulative 
distributions of all 30 systems in our simulated sample.  The dark red envelope encloses the central-most two thirds of these distributions.  
The cumulative distributions of satellite galaxies within 300 kpc of the MW and M31 are also shown, calculated with data from 
\citet{McConnachie2012}.  Middle: Ensemble of  cumulative distributions of satellite stellar masses.  Satellite stellar masses are taken to be 
the total stellar mass within $2 \times r_{1/2}$. Shaded regions have the same meaning as the left panel and data for the MW and M31 are 
also shown.  Right: Ensemble of cumulative distributions of satellite subhalo masses.  The satellite subhalo masses are taken to be the 
total subhalo mass returned by SUBFIND for each system.  The red contours enclose the cumulative distribution of all subhaloes with a 
stellar mass greater than $5 \times 10^5$ \Msun, which is approximately the mass of 10 star particles.  The grey contours enclose the 
cumulative distributions of all subhaloes, including subhaloes without stellar particles.}
\label{fig:cum_distr}
\end{figure*}

Figure \ref{fig:cum_distr} shows the range of mass and luminosity distributions for subhaloes within 300 kpc of the host and how 
this range compares to the satellite galaxy luminosity distributions found in the MW and M31 \citep{McConnachie2012}.  Note that our stellar mass limit is 
significantly above the stellar masses of recently discovered MW satellite galaxies found in the Dark Energy Survey \citep{Koposov2015,DrlicaWagner2015}.  
The MW's satellite luminosity distribution falls within the range of the distributions produced by the thirty Auriga zoom haloes, although due to incomplete sky 
coverage of galactic surveys such as SDSS, the MW distribution is likely incomplete at the fainter end.  The distribution of M31 satellites 
lies at the upper number end of the simulated range.  There is also a large variation in the mass and luminosity of the most massive 
subhalo within each host.

The total number of subhaloes within 300 kpc that have a total subhalo mass above $10^8$ \Msun\ and a stellar mass above 
$5 \times 10^5$ \Msun\ across all thirty host halo zooms is 356; within 1 Mpc, the total is 673.  There is a small population of systems that lie above the stellar 
mass cut but below the total halo mass cut: there are 31 of systems within 300 kpc, and within 1 Mpc, there are 35.  There is 
significant variation in the number of subhaloes between simulated hosts.  Within 300 kpc the number ranges from just 6 subhaloes (Au22) 
to 18 subhaloes (Au15).  The average number of selected subhaloes within 300 kpc is 11.9  Again, these numbers apply to subhaloes 
meeting our stellar mass and total mass cuts.  

Most of the subhaloes that meet the total mass cut of $10^8$ \Msun\ do not meet the stellar mass cut of $5 \times 10^5$ and have either 
lower stellar masses or are completely dark.  These systems are shown in the right most panel in Figure \ref{fig:cum_distr}. 
Within 300 kpc, only 20\% of subhaloes above $10^8$ \Msun\ in total mass meet our stellar mass threshold and within 1 Mpc, only 10\% do so.  Many of the 
systems excluded by our selection criteria are completely dark.  Within 300 kpc, 70\%  of all subhaloes with masses above $10^8$ \Msun\ are dark and within 
1 Mpc 84\% are dark.  The most massive entirely dark system has a subhalo mass of $5.4 \times 10^9$ \Msun\ and all subhalos with 
subhalo masses above $5.6 \times 10^9$ \Msun\ have stellar masses above our stellar mass threshold of $5 \times 10^5$ \Msun.  These results, along with 
the subhalo mass and luminosity distributions of more luminous satellite systems shown in Figure \ref{fig:cum_distr}, are consistent with the findings of the 
APOSTLE simulations that used a different model for subgrid star formation and feedback physics and for hydrodynamics \citep{Sawala2016b}.

\subsection{Gas properties}

We explore the gas content of subhaloes by estimating the mass of neutral hydrogen gas within each subhalo system.  
Following the methods of \citet{Marinacci2017}, we compute the total H~I mass within each gas cell.  For non-star forming cells, this mass 
comes directly from the fraction of neutral hydrogen given by the atomic rate equations that are solved to give non-equilibrium atomic cooling rates.  
For star-forming cells, we apply assumptions of the two-phase subgrid ISM model employed in the Auriga simulations, where gas in star-forming cells is 
assumed to be split between a hot phase and a cold phase that are in pressure equilibrium and is empirically calibrated to observations of nearby 
galaxies \citep{Leroy2008,BlitzRosolowsky2006}.

Using these H~I cell masses, we compute for each subhalo a spherically averaged, radial volume density profile for the H~I gas.  This spherical volume 
density profile is then projected with the Abel transform integral to compute a radial H~I column density profile for the subhalo.  From this H~I column density 
profile, the radius where the H~I column density drops below $10^{20}$ cm$^{-2}$ is found.  The total H~I mass for the subhalo is then taken to be the sum of 
H~I cell masses of all gas cells within this computed radius.  The column density $10^{20}$ cm$^{-2}$ is $\sim$10\% of the peak column density of Leo~T, 
one of the lowest luminosity Local Group dwarf galaxies to contain H~I \citep{RyanWeber2008}.  We found that H~I system masses did not substantially 
increase for threshold values lower than this value.

\begin{figure*}
\centering
\includegraphics[width=\textwidth]{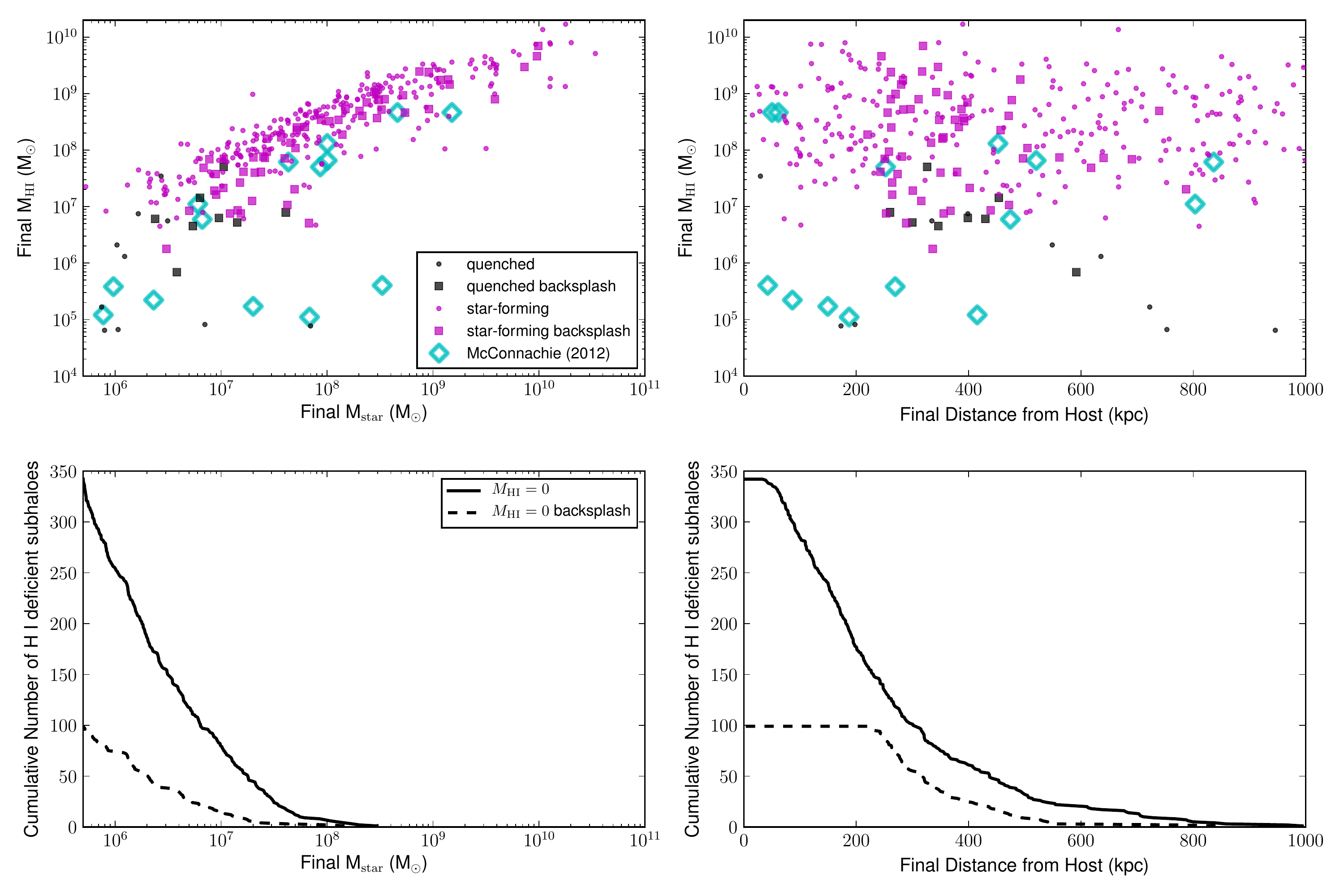}
\caption{H~I masses of subhaloes within 1 Mpc versus final stellar mass (top left) and final distance from the host (top right).  All systems with non-zero H~I 
masses are shown.  Quenched systems are shown with back points and star-forming systems are shown with magenta points.  `Backsplash' systems 
(systems beyond \R\ at $z=0$ that have passed within \R\ at some earlier redshift) are shown with squares; satellites and non-backsplash isolated systems 
are shown with circles.  Measured H~I masses of Local Group systems within 1 Mpc of either the MW or M31  
\citep{McConnachie2012} are also shown (cyan diamonds).  Systems that have zero H~I masses are shown in the lower panels: cumulative distributions of 
the number of H~I deficient systems with stellar masses greater than plotted value (left) and the number of H~I deficient systems with final distances greater 
than the plotted value (right).  Distributions for H~I deficient backsplash systems are shown with dashed lines.  
Most H~I systems have lower stellar masses and/or are closer to their host. }
\label{fig:HIscatter}
\end{figure*}

\begin{figure}
\centering
\includegraphics[width=\columnwidth]{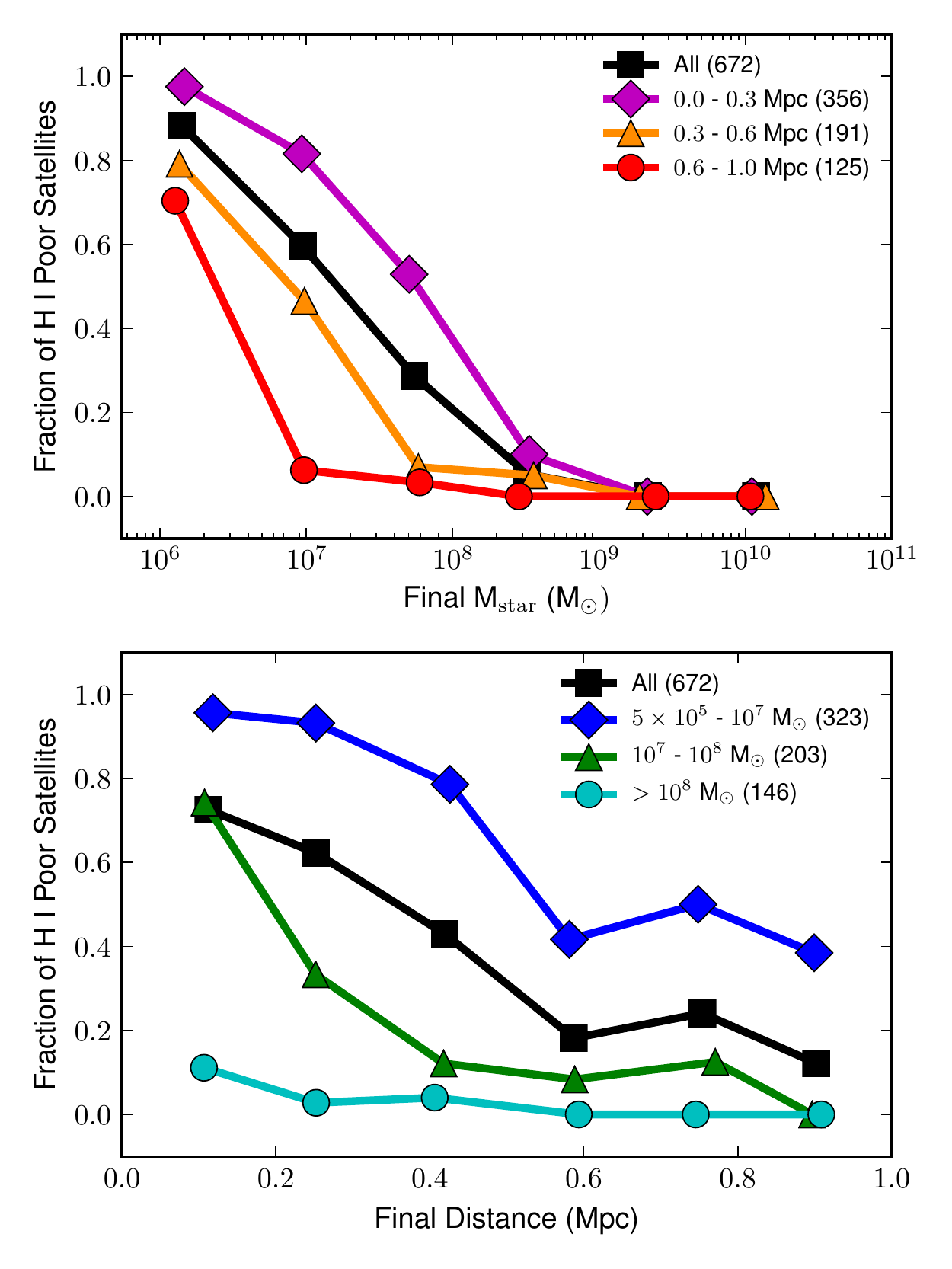}
\caption{Top: Fraction of subhaloes with H~I masses below $10^5$ \Msun\ (called here `H~I Poor') plotted versus final stellar mass.  
Trends for subhaloes found in different distance ranges are plotted with different colors and symbols as labeled and the total number of 
subhaloes in each distance trend is indicated in parentheses after each label.  Bottom: Fraction of H~I poor systems plotted versus final subhalo-host 
distance.  Trends for subhaloes with different final stellar mass ranges are plotted with 
different colors and symbols as labeled and the total number of systems in each mass range is indicated in parentheses after each label.  }
\label{fig:HIpoor}
\end{figure}

Figure \ref{fig:HIscatter} shows the H~I mass of subhaloes versus their stellar mass and distance from their host. There appears to be a correlation between 
stellar mass and H~I mass for H~I rich systems.  There does appear to be a strong environmental effect as well, as shown in Figure \ref{fig:HIpoor}, especially 
for subhaloes with stellar masses below $10^8$ \Msun.  This is consistent with the overall environmental trend found for the H~I content of MW and M31 
satellites \citep{Grcevich2009}.  Figure \ref{fig:HIscatter} also shows that a significant fraction of isolated systems with zero H~I mass are `backsplash' 
systems, which we will discuss in Section \ref{sec:backsplash}.  It does appear, however, that our estimate of H~I mass is a factor of a few greater than in 
observed H~I rich systems at the same stellar mass.  It is possible that this overestimate is due to unaccounted for ionizing sources (such as massive stars 
within the subhaloes themselves) or limitations of our ISM subgrid model.  Among systems with a non-zero H~I mass, there does not appear to be a 
correlation between final host distance and H~I mass.

Despite these limitations, our estimates of the H~I content do appear to be robust enough to examine general trends.  We 
separate subhaloes into two groups: `H~I rich' systems that have an H~I mass greater than $10^5$ \Msun\ and `H~I poor' systems that 
have an H~I mass less than $10^5$ \Msun.  Most of the H~I poor systems in fact have little to no gas mass associated with them.  
Figure \ref{fig:HIpoor} shows the fraction of H~I poor systems across the sample.  For systems with stellar masses below $10^7$ \Msun, 
roughly the stellar mass of the Fornax satellite galaxy, over 80\% are H~I poor within 300 kpc of the host.  There appears to be a sharp 
drop in the fraction of H~I poor systems around 500 kpc.  Beyond 500 kpc, less than 40\% of systems below $10^7$ in stellar mass are H~I 
poor.  

In all distance bins, there is a strong stellar mass dependence for the H~I content.  Very few systems with a stellar masses above $10^8$ 
\Msun\ (roughly the stellar mass of the SMC) are H~I poor.  The stellar mass at which virtually no system is H~I poor drops for larger 
distance bins; for example, systems that are more than 600 kpc from the host are mostly H~I rich down to a stellar mass of $10^7$ \Msun.

\subsection{Star-formation properties}
\label{sec:sfproperties}
\begin{figure}
\centering
\includegraphics[width=\columnwidth]{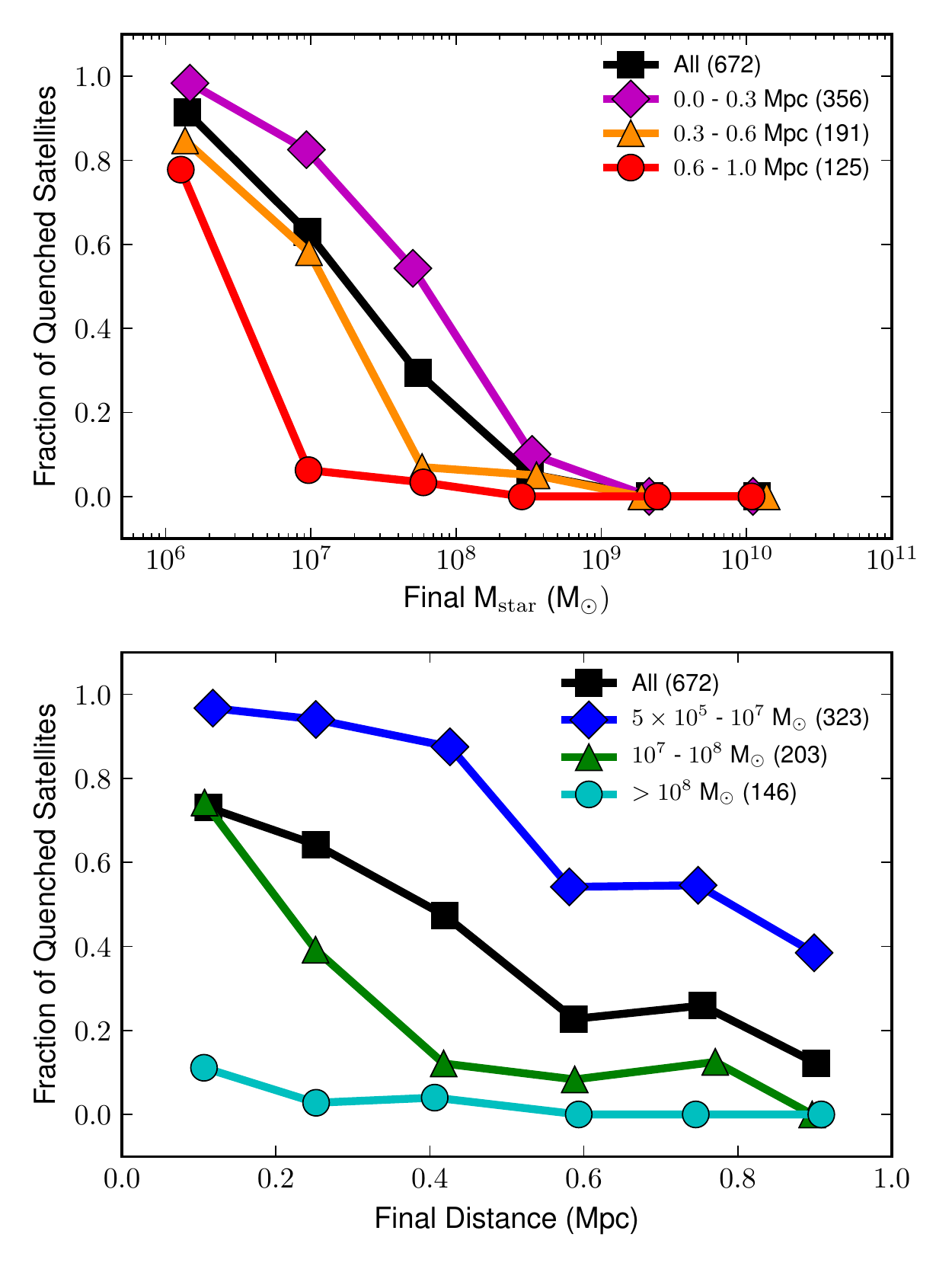}
\caption{Top: Fraction of quenched subhaloes versus final subhalo stellar mass at redshift zero.  Quenched subhaloes are defined as those 
whose youngest star particle is more than 100 Myr old and whose final gas phase SFR is zero.  Trends for different distance bins are 
shown with the same bin ranges and colors as Figure \ref{fig:HIpoor}.  The total number of satellites in each trend is labeled.  
Bottom: Fraction of quenched subhaloes versus final host-subhalo distance.  Trends for different stellar mass bins are shown with the same 
bin ranges and colors as Figure \ref{fig:HIpoor}.  The total number of satellites in each trend is labeled.}
\label{fig:qfrac}
\end{figure}

The stellar mass and environmental trends seen in the H~I content of subhaloes are also seen in the star-forming properties of subhaloes.
Figure \ref{fig:qfrac} shows these trends, with regards to the fraction of quenched subhaloes.  Here, and in subsequent analysis, we define 
a `quenched' subhalo as being one whose youngest star particle is more than 100 Myr old and whose $z=0$ star formation rate (SFR) 
is zero.  A subhalo's SFR is computed as the sum of the SFRs of individual gas cells linked to the system by the SUBFIND algorithm. 

Across the entire sample, no subhalo with a stellar mass of $10^9$ \Msun\ or greater is quenched, regardless of environment, and most 
subhaloes with a stellar mass above $10^8$ \Msun\ are also unquenched.  The majority of low-mass systems ($M_{\rm{star}} < 10^7$ 
\Msun) are  quenched.  At these masses, the fraction of quenched systems tends to be higher than the fraction of H~I poor systems.  At a 
stellar mass of $10^6$ \Msun, the fraction of quenched systems is over 70\% regardless of distance from the host.  For the most distant 
low-mass systems (beyond 500 kpc), less than 60\% are quenched, compared to less than 40\% being H~I poor.  This indicates a 
population of quenched subhaloes with H~I masses greater than $10^5$ \Msun.

\section{Quenching Processes}
\label{sec:processes}

\begin{figure*}
\centering
\includegraphics[width=\textwidth]{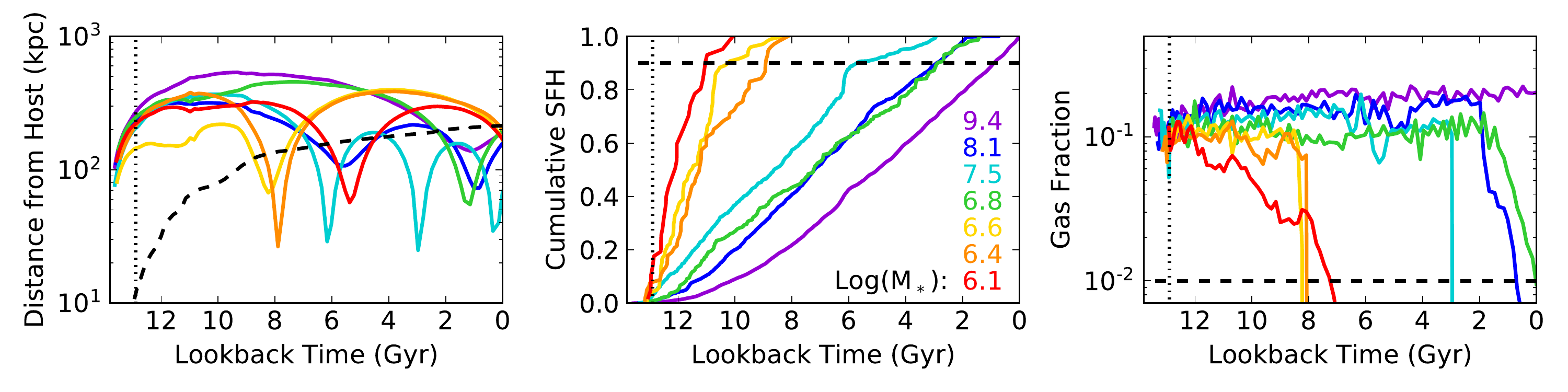}
\caption{Evolution of luminous satellites of Au6.  The plotted satellites all end the simulation within $R_{200}$, have a minimum final subhalo 
mass of $10^8$ \Msun, and a minimum stellar mass within $2 \times r_{1/2}$ of $5 \times 10^5$ \Msun.  The vertical dotted line in each panel marks $z=6$, 
the time when reionization concludes.  Left: Distance in physical kpc 
between the host halo and the satellite halo versus lookback time.  Each subhalo is plotted as a differently colored line; the $R_{200}$ 
radius of the host halo (which evolves with time) is plotted with a black dashed line.  Middle: Cumulative star formation history of star 
particles that end the simulation within $2 \times r_{1/2}$ versus lookback time.  For reference, a threshold value of 0.9 is indicated with a 
black dashed line.  The logarithm of the final stellar mass of each system is indicated in the corresponding color.  Right: Satellite gas 
fraction (gas mass divided by the total mass) within $2 \times r_{1/2}$ vs. lookback time.  For reference, a threshold value of 0.01 is indicated with a black 
dashed line.}
\label{fig:halo_evo}
\end{figure*}

We now discuss the physical mechanisms driving quenching in subhaloes.  
The evolution of the gas content of individual subhaloes is suggestive of some quenching mechanisms, such as ram pressure stripping.
Figure \ref{fig:halo_evo} presents the cumulative star formation histories and gas fraction evolution of several subhaloes that end the 
simulation within \R\ of one of Auriga hosts.  
The example systems quench at a variety of lookback times ranging from more than 10 
Gyrs ago to less than two, and one system remains star-forming and gas rich at redshift zero.  While star-forming, most satellites maintain 
gas fractions close to 0.1.  
When they lose gas, many systems do so rapidly, and once lost, none of these example systems recover their gas.  
This rapid drop in the central gas fraction may be a signature of ram pressure stripping of the subhalo gas.  

The spatial distribution of subhaloes beyond \R\ (Figure \ref{fig:host_image}) is also suggestive of larger-scale environmental effects.  Many subhaloes 
appear to be accreted into the host halo environment from large scale filaments that have their own associated gas reservoirs and contain many other 
subhaloes, which may both play a role in the baryon histories of systems.  The impact of feedback in our model should also be considered 
and effects from the cosmic UV background may also play a role.

\begin{figure*}
\centering
\includegraphics[width=\textwidth]{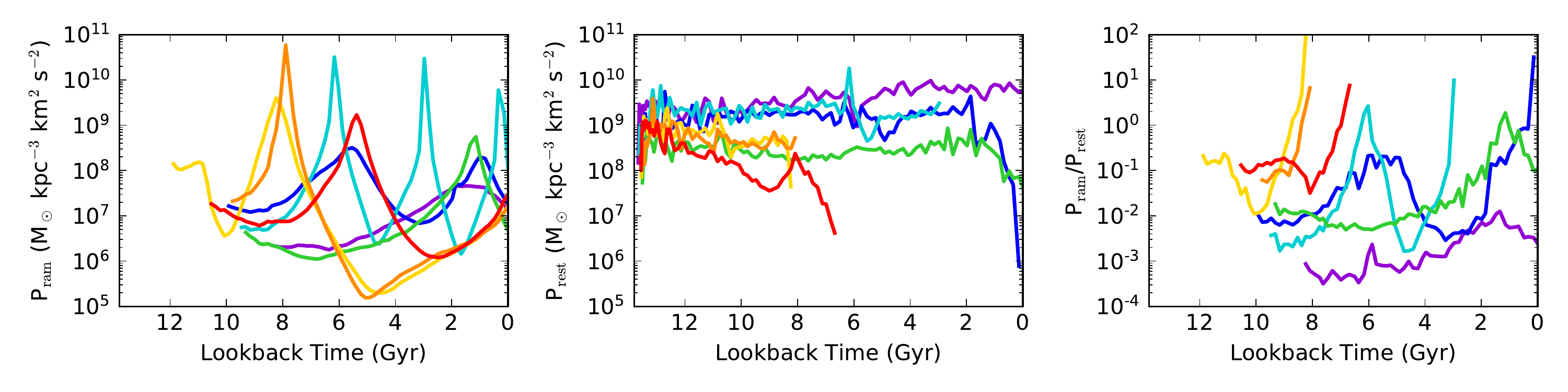}
\caption{Evolution of pressure forces felt by gas in the luminous satellites of Au6.  Satellites plotted end the simulation within \R\ and are 
the same as those plotted in Figure \ref{fig:halo_evo} with the same colors.  Left: Estimate of the ram pressure felt by satellites versus 
lookback time as computed from a time-dependent average gas density profile of the host and the satellites own velocity.  Each line begins 
when the satellite comes within 4\R\ of the host.  Middle: Estimate of the restoring force per area felt by gas within each satellite.  Each line 
begins when the satellite first forms stars and ends once the satellite's gas mass falls to zero.  Note that following Newton's gravitational law, the force 
between a subhalo's stars and gas is only non-zero when the mass of both stars and gas within the subhalo is non-zero.  Right:  Ratio of the ram pressure to 
the restoring force per area plotted in the first two panels.  Each line begins and ends when either $P_{\rm{ram}}$ or $P_{\rm{rest}}$ is undefined.}
\label{fig:halo_rampressure}
\end{figure*}

\subsection{Ram Pressure Stripping}
\label{sec:rampressure}

We begin with an examination of ram pressure stripping to determine whether it is responsible for the rapid drops in gas content that many 
subhaloes exhibit when they are accreted into the host halo as seen in this study and others \citep[e.g.][]{Okamoto2009}.  Ram pressure stripping 
refers to the phenomenon that occurs when the pressure exerted on a galaxy's internal gas by its motion through an ambient medium (such as the CGM or 
the ISM) overcomes the gravity exerted by stars and dark matter within the galaxy on the gas \citep{GunnGott}, resulting in a stripping of the internal gas from 
the galaxy in motion.  

The ram pressure felt by a satellite's gas is $P_{\rm{ram}} = \rho_{\rm{CGM}} v^{2}_{\rm{sat}}$, where $\rho_{\rm{CGM}}$ is the density of 
the medium though which the satellite galaxy is moving and $v_{\rm{sat}}$ is the relative velocity of the satellite to the surrounding gas.  
The restoring force per area on the satellite's gas can be expressed as 
$P_{\rm{rest}} = \left | \frac{\partial \Phi}{\partial z_h} \right |_{\rm{max}} \Sigma_{\rm{gas}} $, 
where $\Sigma_{\rm{gas}}$ is the satellite's gas surface density, $z_h$ is the direction of motion (and gas displacement), $\Phi$ is the gravitational potential, 
and $\left | \frac{\partial \Phi}{\partial z_h} \right |_{\rm{max}}$ the the maximum of the derivative of $\Phi$ along $z_h$ \citep{Roediger2005}.

Using the merger trees, we have estimated for each satellite the ram pressure and restoring force felt by its internal gas over its history. For 
the quantity $v_{\rm{sat}}$, we adopt the velocity of the subhalo relative to the host halo's velocity.  For the quantity $\rho_{\rm{CGM}}$, we 
compute an average radial gas density profile extending out to a radius of 4\R\ for the host halo in each snapshot.  The quantity 
$\rho_{\rm{CGM}}$ is then interpolated from this average gas density profile in the corresponding snapshot at the subhalo's radial position.  
The average radial gas density profile is computed over all gas cells belonging to the host's primary subhalo and unlinked cells (cells that do not belong to any 
halo or subhalo).  The values for $P_{\rm{ram}}$ that we compute from these two values will in many cases not capture the maximum ram pressure felt at 
pericentric passages because the spacing of our outputs is not sufficient to capture the true pericentric distance for many orbits.

Satellites in Auriga encounter many local perturbations in the density and velocity fields of the gas that they encounter along their orbital path.  These 
variations arise from feedback-driven winds from the host galaxy, clumpy gas accretion from the IGM, and stripped gas from other satellites.  Our strategy for 
estimating the ram pressure felt by satellites neglects these local variations, however, with the refinement strategy adopted in the Auriga simulations, gas at 
lower densities (as found in the outskirts of the host haloes) is resolved with fewer resolution elements.  This is a result of the simulations' Lagrangian 
refinement strategy that ensures an approximately constant gas cell mass throughout the simulation.  The median cell diameter of star-forming gas in the host 
disks is approximately 300 pc, but beyond 10\% of \R\ cells typically have a diameter greater than 1 kpc.  An estimate of the local ram pressure would, at large 
distances, involve averages over small numbers of cells and be sensitive to computational choices.  Our radially averaged estimate, while neglecting local 
effects, robustly captures the effect of radial infall that drives the main change in ram pressure, which can vary by orders of magnitude.

The restoring force estimates often assume a disk configuration for the galaxy within a spherical halo \citep{GunnGott,Roediger2005}, which is the case for 
some systems in our sample, but not the case for others.  We therefore adopt a simple estimate for $\Sigma_{\rm{gas}}$ and 
 $\left | \frac{\partial \Phi}{\partial z_h} \right |_{\rm{max}}$ that can be applied to all systems in the sample uniformly.  The gas surface density is estimated 
 from the radius enclosing half the gas mass ($r^{\rm{gas}}_{1/2}$), 
such that $\Sigma_{\rm{gas}} = M_{\rm{gas}}/ 2\pi (r^{\rm{gas}}_{1/2})^2$ (where $M_{\rm{gas}}$ is the total mass in gas).  We estimate  
$\left | \frac{\partial \Phi}{\partial z_h} \right |_{\rm{max}} \sim v^2_{\rm{max}}/r_{\rm{max}}$, where $v_{\rm{max}}$ is the maximum velocity of the 
spherically-averaged subhalo rotation curve, and $r_{\rm{max}}$ is the radius where this peak occurs.

Our simulations' ability to capture the restoring force depends in part on the accuracy of our gravity calculation.  The minimum softening length used is more 
than 300 pc in these simulations and the typical half-light radius of observed systems at our lower luminosity limit is close to this value.  (We also note that gas 
cells can and do have radii smaller than the minimum softening length in dense gas.)  We find typical half stellar mass spherical radii of approximately 1-2 kpc, 
however, we have not attempted to model properly the shapes of the satellite systems, which is typically done for observed systems.  Many of our systems are 
very squashed or have stellar disks, and the spherical radii are overestimates for this reason.

When examining the dynamical mass within 300 pc, we find values between $10^6$ \Msun\ and  $10^7$ \Msun\ for systems with stellar masses up to 
$10^7$ \Msun, with our higher resolution simulations being slightly closer to $10^7$ \Msun.  This is consistent with observed systems \citep{Strigari2008}.    
We also find small differences in both these quantities with our high resolution simulations (described in Appendix A) that have a factor of two better softening 
lengths.  We conclude that we are capturing the central potentials of these systems reasonably well for the purposes of this study, but defer a fuller discussion 
of the internal structure of these objects to a study of our higher resolution simulations.

Figure \ref{fig:halo_rampressure} shows the evolution of the ram pressure and restoring force felt by satellites over their histories.  Figure 
\ref{fig:halo_rampressure} can be compared to Figure \ref{fig:halo_evo}, which depicts the same satellites, and it can be seen that sharp 
increases in the ratio $P_{\rm{ram}}/P_{\rm{rest}}$, rising above 1, correspond in most cases to a sudden, sharp drop in the gas fraction.

\begin{figure*}
\centering
\includegraphics[scale=0.95]{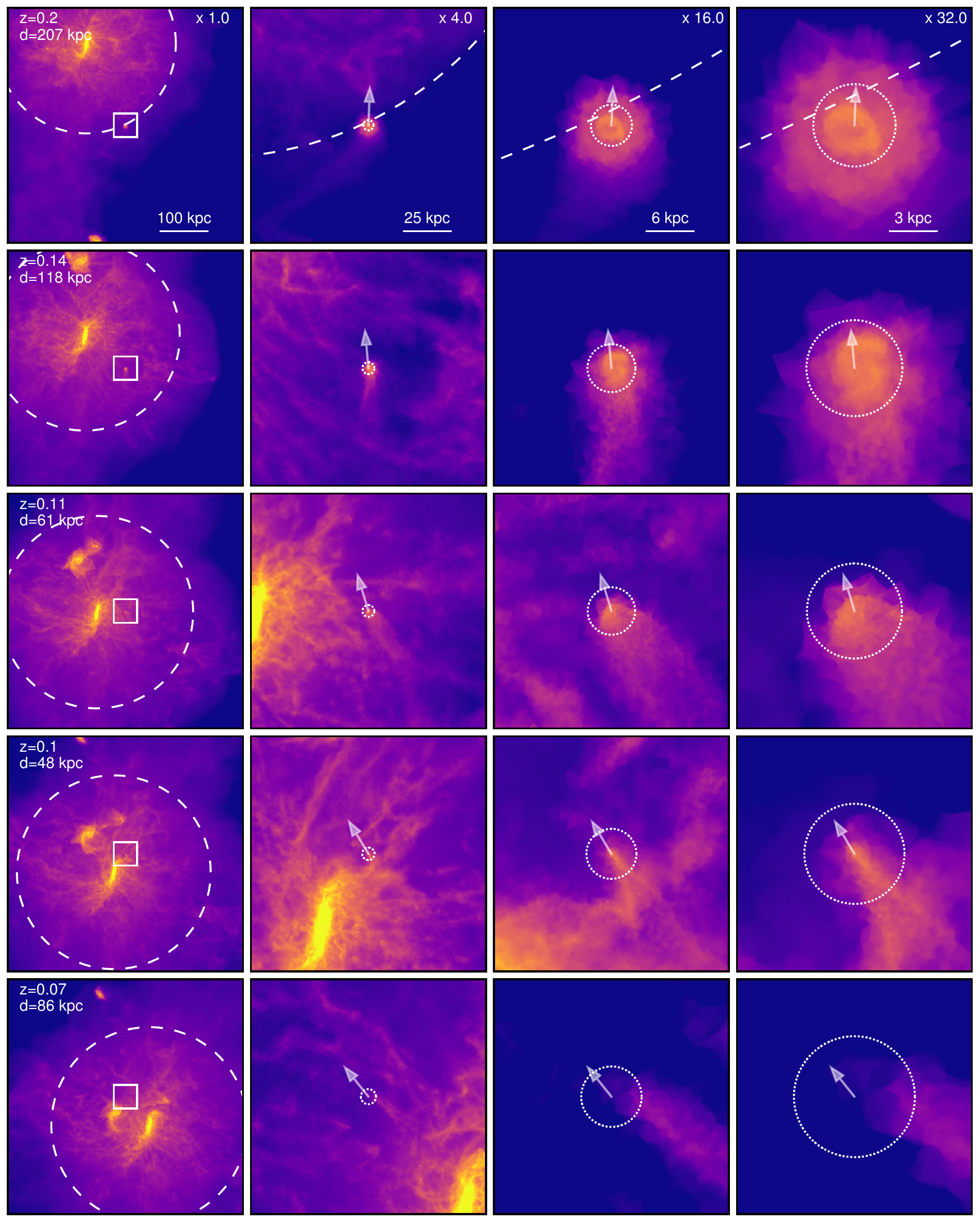}
\caption{Gas density projections of an individual subhalo as it falls into the host halo and is ram pressure stripped.  Each row shows the 
subhalo at a single time and each column shows projections through cubes of decreasing volume from left to right.  The projection cube 
length decreases by the factor indicated for each column and are, from left to right, 500 kpc, 125 kpc, 31 kpc, and 16 kpc.  In the left 
most column, the subhalo is surrounded by a 50 kpc square.  In the other three columns, the subhalo is surrounded by a 
dotted-white circle that has a radius equal to twice the half-stellar mass radius.  The direction of the subhalo's velocity in the plane of the image 
is shown with an arrow.  The host halo's $R_{200}$ radius is indicated by a dashed white circle.  The redshift and 
the host-subhalo distance are indicated in each row.  These images show an example subhalo from a resimulation of Au6 that was run with 
8 times more snapshots.  The subhalo does not reaccrete gas after this stripping event and ends the simulation composed only of 
stars and dark matter.}
\label{fig:stripping_image}
\end{figure*}

There are a few interesting cases among this group of satellites.  The system plotted in dark blue in Figures \ref{fig:halo_evo} and 
\ref{fig:halo_rampressure} appears to have a sharp drop in its gas fraction that does not quite match in time its sudden increase in the 
$P_{\rm{ram}}/P_{\rm{rest}}$ ratio.  A closer examination of this system revealed that the drop in the gas fraction is due to an interaction 
with another satellite (in fact, the purple satellite) that is more massive and gas rich.  The gas loss does appear to be due to ram pressure, 
but instead from the gas disk of this secondary satellite rather than from the host's gas.  Our estimate for $P_{\rm{ram}}$ from an average 
gas density profile is inaccurate in a case like this.

The system plotted in light blue has a complicated orbital history that includes three pericentric passages.  All three of these pericentric 
passages correspond to a spike in the ram pressure, but during the first pericentric passage, the restoring force of the system is great 
enough to counter the ram pressure.  What occurs instead is a compression of gas within the satellite resulting in a small star burst 
(apparent in the cumulative SFH in Figure \ref{fig:halo_evo}).  However, the result of this star burst is to lower the restoring force of the 
satellite (perhaps due to feedback effects associated with the star burst).  On the subsequent pericentric passage, the ram pressure is able 
to overcome the now lower restoring force, resulting in a complete stripping of gas from the subhalo.

A visual example from our resimulation of Au6 with an increased number of outputs is shown in Figure \ref{fig:stripping_image}.  Here a 
satellite that ends the simulation within \R\ and with a stellar mass of $6.3 \times 10^6$ \Msun\ is shown during a low-redshift stripping 
event.  Even before this system crosses \R, it appears to already have a low density gas tail.  As the system moves toward pericentre, a 
more pronounced gas tail forms and the gas within the centre of the halo appears to form a contact discontinuity surrounded by a weak 
bow shock.  The edges of this structure are eroded and a small bullet-like clump of gas remains, until it too is eroded, and all the 
gas is stripped from the subhalo.  This system's stellar mass does not change appreciably during this event.

\begin{figure}
\centering
\includegraphics[width=\columnwidth]{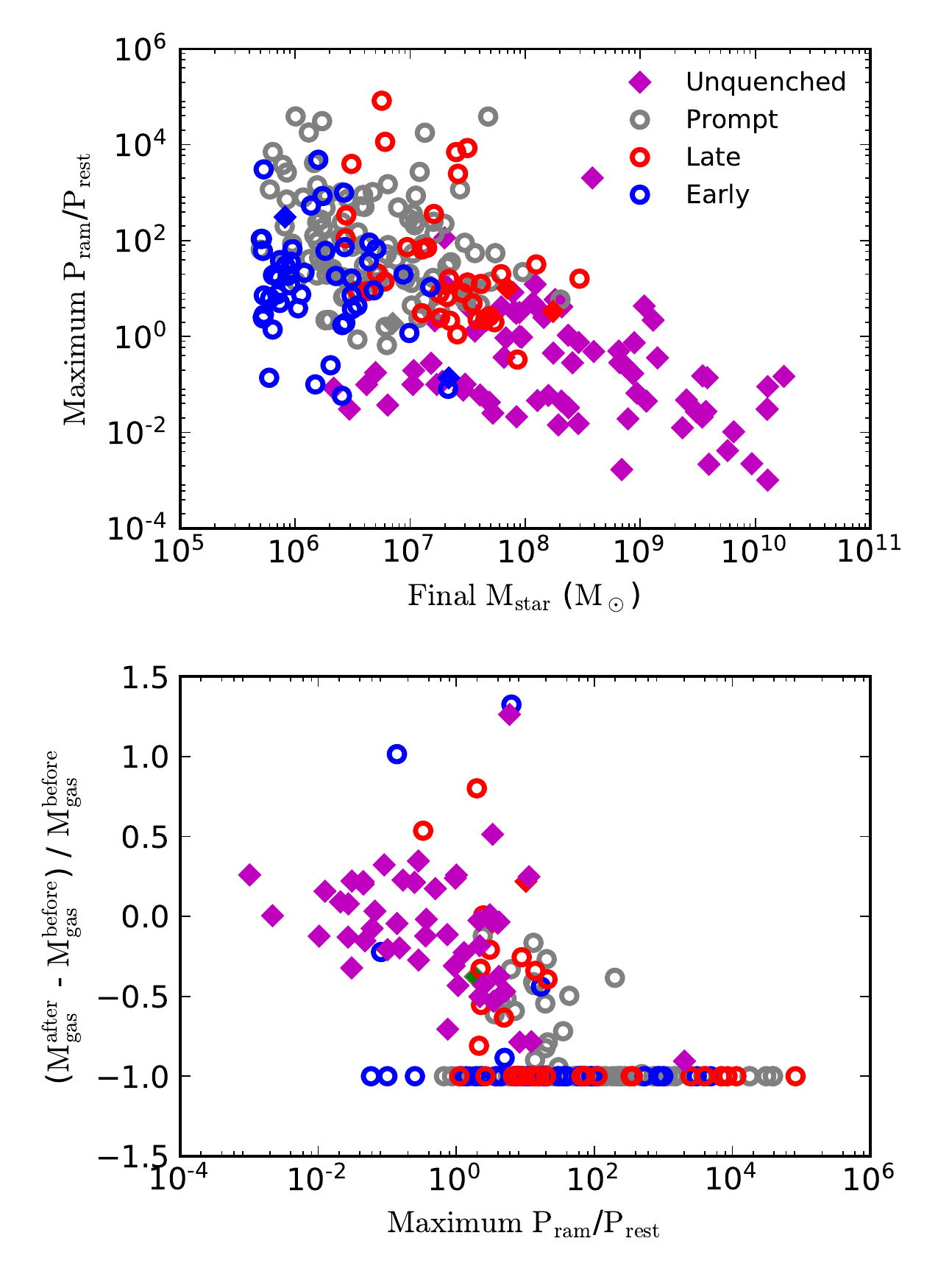}
\caption{Top: The peak value in the ratio $P_{\rm{ram}}/P_{\rm{rest}}$ felt by each satellite system that ends the simulation within \R\ of its 
host.  Systems where the maximum value occurs in the final snapshot are not included as well as systems where $P_{\rm{ram}}/
P_{\rm{rest}}$ is undefined (mainly systems that lose their gas prior to coming within 4\R\ of the host).  Open circles show systems with no 
gas and solid diamonds show systems with a non-zero final gas mass.  Magenta points are unquenched systems; other colors indicate the 
system's membership in the quenching samples 
described in Section \ref{sec:satevo}.  Bottom: The fractional change in gas mass of satellite systems between the snapshot directly before 
the peak $P_{\rm{ram}}/P_{\rm{rest}}$ ratio occurs and the snapshot directly after.  Symbol styles and colors have the same meaning as 
the top panel. }
\label{fig:ramfrac}
\end{figure}

In Figure \ref{fig:ramfrac}, we present the effect of ram pressure on the satellite sample as a whole for systems within \R\ at $z=0$.  Here 
we identify the maximum in the ratio $P_{\rm{ram}}/P_{\rm{rest}}$ over the history of each satellite.  We examine the correlation of this ratio 
with the final stellar mass of the system and the change in gas mass at the time of the ratio peak.   The peaks in 
$P_{\rm{ram}}$ are likely lower limits to the true peak values because of the cadence in our outputs and the very steep rate of change in 
this value.  

A majority of unquenched systems do not reach values in $P_{\rm{ram}}/P_{\rm{rest}}$ above 1 ($\sim$29\% do so) and most do not reach 
values above 10 (only $\sim$5\%).  Among quenched systems, most reach $P_{\rm{ram}}/P_{\rm{rest}}$ ratios above 1 ($\sim$95\%) and
the fractional change in gas mass between the snapshot before the peak $P_{\rm{ram}}/P_{\rm{rest}}$ and the snapshot after, is nearly 
100\% for over 70\% of quenched systems.  What this gas loss means for the star formation histories of these systems likely varies; for 
systems that quench before infall, they may have already been quenched, and for the systems that form stars after this gas stripping event 
occurs, they may reaccrete gas and have later epochs of star formation.
 
\begin{figure}
\centering
\includegraphics[width=\columnwidth]{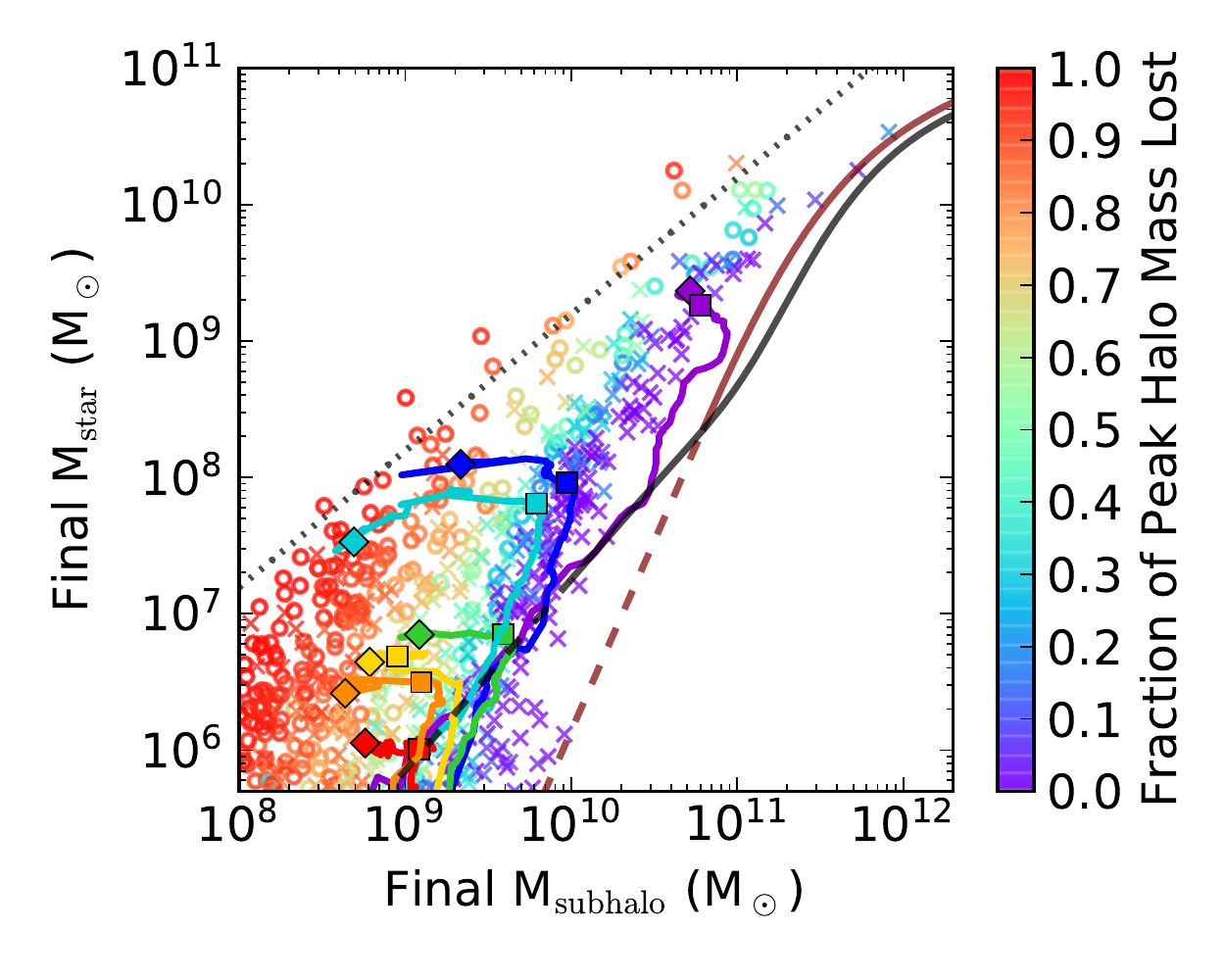}
\caption{The final stellar mass versus final halo mass for all resolved systems within 1 Mpc of the host galaxies in all 30 Auriga simulations.  Open circles are 
systems within \R\ at $z=0$ and x's are systems beyond \R.  The color of these points indicate the fraction of their total halo mass that has been lost from their 
peak mass as measured from their merger trees: red points have lost over 90\% of their halo mass and purple points have lost very little mass.  Colored lines 
show the evolutionary tracks of a few example systems from Au6 in this plane.  Along these tracks, squares indicate the point at which the system first crosses 
within \R\ and diamonds indicate their final position.  The black dotted line indicates the cosmic baryon fraction $\Omega_b / \Omega_m$.  The brown solid/
dashed line indicates an abundance matching estimate for central dark matter haloes taken from \citet{Guo2010} and the black solid/dashed line indicates an 
estimate from \citet{Behroozi2013}.  For both these trends, the solid portion of the line covers the portion of parameter space where these studies make 
predictions and the dashed lines are power-law extensions of the estimated trends to lower masses.   }
\label{fig:mstarmhalo}
\end{figure}
 
\subsection{Tidal effects and starvation} 
As we will discuss in later sections, about half of quenched systems in our satellite sample quench either significantly before or significantly 
after first falling into their host halo, indicating a role for quenching mechanisms other than ram pressure stripping.  Most of the systems that 
quench after infall are higher mass.  High-mass systems that quench significantly after first infall may be thought of as an extension of the 
unquenched population that have simply had enough time to cease star formation given the longer quenching time needed in higher-mass 
systems.  Indeed, all of the late, high-mass quenchers appear to have had an interaction with the host halo prior to 6 Gyr ago.

Investigating some of the high-mass, late quenchers reveals that a number of them show a slow down in their star formation rate after accretion into the host.  
These systems are few in number, so it is difficult to determine population wide statistics.  They appear to have orbits with large angular momenta that prevent 
close pericentric passages, but do in the end undergo a final ram pressure stripping event after a long period of star formation slowdown.  When the slowdown 
occurs, the total gas depletion time (the gas mass divided by the star formation rate) in these systems actually goes up because the star formation rate drops 
faster than the gas mass.  This is somewhat different from the classical starvation quenching picture, but may be due to our model's ISM treatment.

Tidal effects may play a role in quenching these systems; however, these effects would need to act faster than ram pressure, but at the same time, be slow 
enough to produce quenching before complete destruction of the subhalo.  Higher-mass systems that are able to resist the effects of ram pressure for a 
longer period of time are candidates for this process.  Figure \ref{fig:mstarmhalo} shows that many systems in our sample have lost a substantial fraction 
($> 80$\%) of their peak halo mass by $z=0$ and have large stellar masses for their halo masses as compared to trends from abundance matching 
studies \citep{Guo2010,Behroozi2013} and the cosmic baryon fraction ($\Omega_b/\Omega_m$).  

Recent idealized simulations by \citet{Safarzadeh2017} suggest that ram pressure stripping and tidal stripping are interrelated processes in low mass, gas rich 
systems; when gas is stripped from gas rich satellites, they find this sudden change in the satellite's central potential can impact orbits of stars within the 
system and enhance tidal stripping.

The evolution of some example systems in Figure \ref{fig:mstarmhalo} shows that the subhalo mass of these systems can drop by several factors before the 
stellar mass is affected. In some more massive systems, there is a modest increase in stellar mass post first infall.  Indeed, there is very little negative change 
seen in the stellar mass of tidally affected satellites.  This is probably due to the high concentration of the stellar component compared to the dark matter, and 
it is likely that once the stellar component of the subhalo is tidally affected, the timescale for the subhalo destruction is short.  The main effect of tides on the 
star formation histories of satellites in our model is therefore most likely the suppression of gas accretion, which will starve systems and will increase the 
effectiveness of other mechanisms, such as ram pressure stripping or stellar feedback.

\subsection{Field quenching and low-mass systems}
The quenching trend with mass for isolated systems demonstrates a sharp increase in quenched fractions below $M_{star} = 10^7$ \Msun.  This indicates a 
field quenching mechanism that is effective in quenching low mass systems.  This mechanism appears to operate in our higher-resolution simulations (see 
Appendix A), which rules out resolution quenching.  It is also notable that systems beyond \R\ that have never had an interaction with a host halo demonstrate 
a very strong mass-dependent quenching signal (as shown in Figure \ref{fig:qfrac_infall}) which indicates that a `backsplash' population cannot be the sole 
explanation.

\begin{figure*}
\centering
\includegraphics[width=\textwidth]{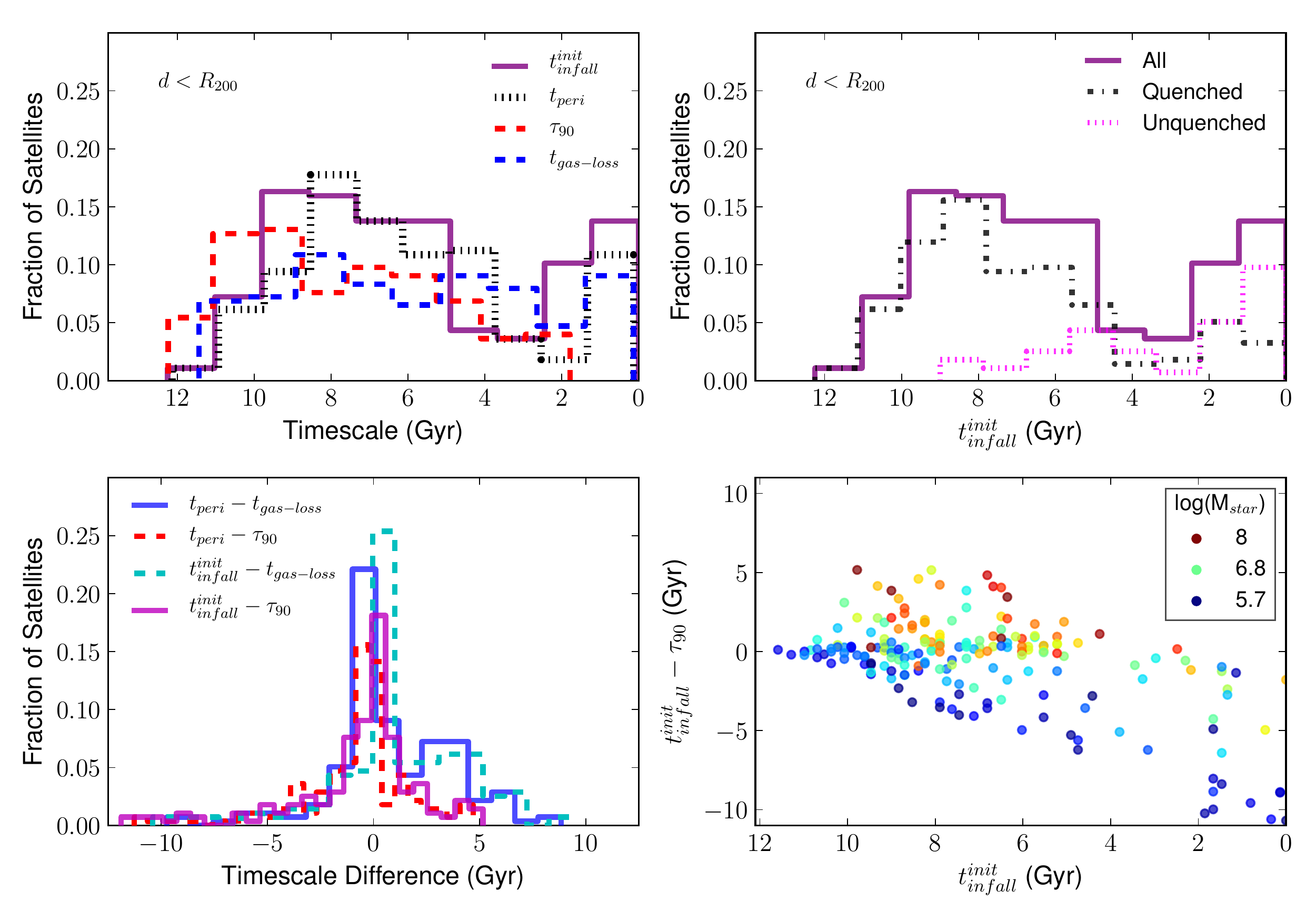}
\caption{Distributions of evolutionary timescales (as described in Section \ref{sec:timescales}) for satellites that lie within \R\ at redshift 
zero in all 30 simulations.  All timescales are measured as lookback times, where 0 is the present day and larger timescales correspond to higher redshift. 
Top Left: Distributions of $t^{\rm{init}}_{\rm{infall}}$, $t_{\rm{peri}}$, $t_{\rm{gas-loss}}$, and $\tau_{90}$.  The 
distributions plotted are the frequency of satellites in each bin divided by the total number of satellites within \R.  
Top Right: Distributions of $t^{\rm{init}}_{\rm{infall}}$ for all systems within \R\ and systems that are quenched and unquenched.  
Bottom Left: Distributions of timescale differences.  Plotted here are the differences ($t_{\rm{peri}} - t_{\rm{gas-loss}}$), ($t_{\rm{peri}} - \tau_{90}$), 
($t^{\rm{init}}_{\rm{infall}} - t_{\rm{gas-loss}}$), and ($t^{\rm{init}}_{\rm{infall}} - \tau_{90}$); the median values of each distribution (in the 
same order) are 0.1 Gyr, -0.3 Gyr,  0.6 Gyr, and 0.0 Gyr.  In distributions involving $\tau_{90}$, only quenched systems are included. 
Negative numbers indicate quenching or gas loss before the orbital timescale under comparison. Bottom Right: The difference 
$t^{\rm{init}}_{\rm{infall}} - \tau_{90}$ plotted vs. $t^{\rm{init}}_{\rm{infall}}$ for quenched systems.  The color of each point indicates its final 
stellar mass: blue points are low mass and red points are high mass. }
\label{fig:timescales}
\end{figure*}

Low mass systems that quench before first infall appear to accrete into the host haloes at a steady rate across the sample (as shown by the satellite-host infall 
times shown in Figure \ref{fig:timescales} and discussed in Section \ref{sec:timescales}).    Each of these systems has a unique history and likely multiple 
quenching processes are at play across the sample, but these results implicate either internal processes like feedback, or global processes, such as 
reionization, both of which operate independently of the host galaxies in Auriga.

Stellar feedback may impact the quenching of dwarf galaxies through the generation of galactic winds that expel gas.  Figure \ref{fig:halo_evo} shows that 
baryon evolution of satellites in our sample is relatively smooth in the absence of external effects.  There are few examples of gas loss and reaccretion across 
the entire satellite sample, which would be a possible signature of feedback effects.  Some small starbursts are apparent, but appear to be correlated with 
pericentric passages.  This lack of burstiness is in contrast to high-resolution simulations of dwarf galaxies that model the ISM self-consistently and the 
launching of galactic winds from local feedback processes \citep[e.g.][]{Fitts2016}.  

The feedback model employed here maintains a steady central gas fraction of 10-20\% (relative to the total mass) in equilibrium with gas accretion, but does 
not appear to expel large amounts of gas from subhaloes. It is possible that a feedback model that includes the dynamical impact of stellar feedback on the 
ISM would result in a more bursty star formation history and a different mass loading for galactic winds in dwarf systems.  This effect could alter feedback-
driven quenching.  It is also possible that feedback driven quenching is more effective at stellar masses below our threshold.

A quenching mechanism commonly appealed to in low luminosity Local Group dwarf galaxies is the effect of reionization.  The filtering halo 
mass for the effect of reionization is often assumed to be around $10^9$ \Msun\ because the virial temperature of haloes at this mass is 
similar to the IGM temperature \citep{Gnedin2000,Hoeft2006,Okamoto2008}.  Studies of cosmologically isolated systems at this halo mass have 
demonstrated the photoevaporation of gas during reionization and the subsequent suppression of post-reionization gas accretion 
\citep{Sawala2010,Simpson2013, Shen2014}.  

The sample of systems considered here spans this filtering halo mass transition, and indeed, Figure \ref{fig:host_image} demonstrates that there are many 
dark subhaloes in our simulations and star formation in many of these systems is likely suppressed by reionization.  In terms of reionization fossils, i.e. 
luminous systems that have been quenched by reionization, the inclusion of a stellar mass threshold in our selection criteria preferentially selected systems 
that assembled early.  Figure \ref{fig:cum_distr} demonstrates that there are many more dark and under-luminous subhaloes (systems with 
$M_{star} < 5 \times 10^5$ \Msun) in our simulations above our halo mass limit.  Most of the systems in our sample already had halo masses above 
$10^8$ \Msun\ at $z=6$, and in fact many lower-mass systems in our sample were of a much higher mass at $z=6$ and subsequently lost dark matter 
mass by means of tidal stripping.  It is likely that the stellar mass of reionization fossils with the baryonic physics model employed in the Auriga simulations 
falls below our stellar mass limit. 

To better understand the impact of stellar feedback and reionization on quenching with the Auriga physical model, higher resolution simulations are needed, 
since in our current simulations, these effects likely only play an important role at stellar masses below our stellar mass threshold of $5 \times 10^5$ \Msun, 
or approximately the mass of 10 star particles.  It is also necessary to have higher snapshot cadence at the epoch of reionization (in our model $z = 6$) than 
our current simulations to capture the dynamics of photoevaporative winds.  Simulations with an alternate, or no, UV background and also be informative in 
determine the impact of reionization \citep{Simpson2013}.

An alternate explanation for early, low-mass quenched systems in our sample is environmental quenching outside the main host.
This type of quenching may be a form of ram pressure quenching, where the ram pressure arises from filament gas \citep{BenitezLlambay2013} 
or gas of other subhaloes, rather than gas associated with the host halo.  This type of ram pressure stripping is not captured by our estimates 
presented in Section \ref{sec:rampressure} for host halo ram pressure quenching.  Visual inspection of some individual systems shows filament 
stripping to be present with a somewhat different signature than host stripping that occurs at pericentic passages.  For 
example, in Figure \ref{fig:halo_evo}, the system plotted in red appears to be a case of cosmic web quenching.  The system in question 
crosses a filament feeding the host transversely and this interaction with the filament gas results in a partial stripping of the subhalo that 
quenches star formation.

Another possible effect is simple starvation.  Figure \ref{fig:halo_evo} shows that star-forming satellites maintain a roughly constant central 
gas fraction while the star formation rate is constant, indicating the replenishment of the gas supply.  Environmental effects that interrupt this 
supply could cause quenching.  The source of this gas supply is likely filamentary gas and therefore, the filament environment could affect the 
SFH of satellites.

\section{When do satellites quench?}
\label{sec:timescales}

\begin{figure}
\centering
\includegraphics[width=\columnwidth]{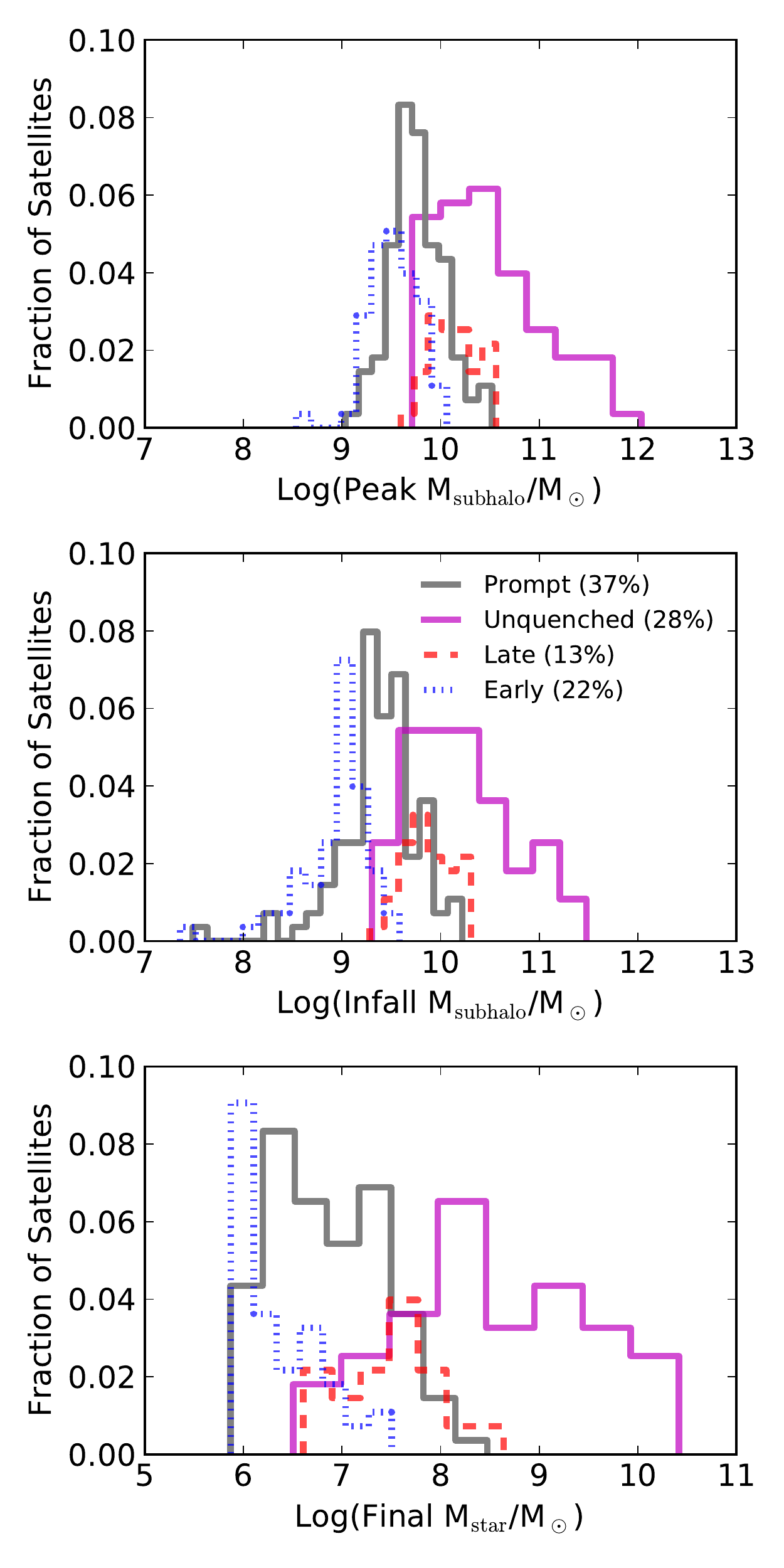}
 \caption{Properties of prompt, early, and late quenching satellites, as well as unquenched satellites, that end the simulation within \R.  The 
percentage of systems in each category is indicated, along with the line style for each distribution.  Top: 
Distribution of the peak subhalo mass in the merger tree history of each subhalo.  The fraction shown is the number 
frequency within each bin divided by the total number of subhaloes (673).  Middle:  Distribution of the subhalo mass at 
first infall ($t^{\rm{init}}_{\rm{infall}}$).  Bottom: Distribution of satellites' final stellar mass.  }
  \label{fig:massesquenched}
\end{figure}
Having established trends in gas content and star formation for subhaloes in our sample at redshift zero in addition to exploring various 
quenching mechanisms, we will now quantify timescales for quenching and gas loss for individual subhaloes and quantify the ensemble of 
these timescales.  We measure timescales on a lookback timescale, where 0 Gyr is the present day and larger times correspond to higher redshift.

\subsection{Timescale Definitions}
First, we define a quenching timescale, $\tau_{90}$, following the observational definition of \citet{Weisz2015}.  All linked star particles that 
end the simulation within twice the satellite's half stellar mass radius ($r_{1/2}$) are used to determine a cumulative star formation history 
for the system using star particles ages.  From this history, $\tau_{90}$ is defined as the lookback time when 90\% of the satellite's final stellar 
mass had been formed.  Note that we will apply this timescale to both quenched and unquenched systems.

Second, we define a timescale for gas loss of gas poor satellites, $t_{\rm{gas-loss}}$.  This timescale relies on the merger trees to track the 
gas content of each system's most massive progenitor at each snapshot backwards in time.  We define the gas fraction as the fraction of 
gas mass to total mass (gas, stars, dark matter, and black holes) within $2r_{1/2}$ and $t_{\rm{gas-loss}}$ as the lookback time when this 
gas fraction drops below 0.01 and remains suppressed below 0.01 down to redshift zero.

Figure \ref{fig:halo_evo} showed the evolution of the baryon content of some example satellite systems.  Also shown in Figure \ref{fig:halo_evo} 
is the distance between the example systems and their host galaxy.  These systems have a variety of orbital histories, with some systems undergoing 
multiple pericentric passages, while others fall in relatively late.  Overall, the baryon histories of these systems appear to be highly correlated with 
their orbital histories, in that the times of star formation quenching, final gas loss, and crossing the host \R\ coincide for many of the systems.

To explore these correlations, we use the merger trees to quantify a timescale for initial satellite infall, $t^{\rm{init}}_{\rm{infall}}$, by 
computing the first time a satellite crosses within the host halo's \R\ radius.  Note that \R\ changes with time.  We also define a timescale for the first 
pericentric passage of each satellite, $t_{\rm{peri}}$, which is the time of the earliest local minimum of the satellite-host distance occurring 
within \R.  Because of discrete time sampling, $t_{\rm{peri}}$ can be over- or underestimated.  If a subhalo never comes within \R, it is not assigned 
a quantity for these timescales.  It is the case, however, that many systems beyond \R\ at $z=0$ have at some point in the past come within \R\ and 
therefore can be assigned values for $t^{\rm{init}}_{\rm{infall}}$ and $t_{\rm{peri}}$.  As can be seen in Figure \ref{fig:halo_evo} many systems have 
complex infall histories and we will explore this in more detail in Section \ref{sec:backsplash}.  

\subsection{Evolution of satellite systems}
\label{sec:satevo}

Figure \ref{fig:timescales} shows the distributions of these timescales for satellites within \R\ and how well they coincide with each 
other for individual systems.  First, it is apparent that the distributions of $t^{\rm{init}}_{\rm{infall}}$ and $t_{\rm{peri}}$ are bi-modal, as is 
$t_{\rm{gas-loss}}$ to a somewhat lesser degree.  The reason for this is discussed in Section \ref{sec:backsplash}.  Second, it is apparent that 
$t_{\rm{gas-loss}}$ and $\tau_{90}$ strongly coincide with 
$t^{\rm{init}}_{\rm{infall}}$ and $t_{\rm{peri}}$.  Of quenched subhaloes, 
51\% quenched within 1 Gyr of first infall and 
48\% quenched within 1 Gyr of their first pericentric passage.  Of gas poor subhaloes, 
43\% lost their gas within 1 Gyr of first infall and 
45\% lost their gas within 1 Gyr of their first pericentric passage.  On average, total gas loss occurs 1.9 Gyr after the cessation of star 
formation as quantified by $t_{\rm{gas-loss}}$ and $\tau_{90}$ for quenched and gas poor subhaloes.

Separating systems by their quenched state, also shown in Figure \ref{fig:timescales}, shows that on average, quenched systems 
tend to have earlier infall times.  Unquenched systems, while less numerous overall, comprise a larger portion of the late-infall systems.

There are of course many quenched systems, 49\%, that do not quench within 1 Gyr of first infall.  Figure \ref{fig:timescales} shows that 
systems that quench before first infall tend to be lower mass and systems that quench after first infall tend to be higher mass.  

Examining Figure \ref{fig:halo_evo}, a few examples of these types are apparent.  A lower luminosity system plotted in red, appears to have 
a rapid cessation of star formation that occurs several Gyrs before its first infall into the host and then a more gradual loss of its gas mass 
that also appears to occur before its first pericentric passage.  There is a system plotted in dark blue that appears to quench over 2 Gyrs 
after its first infall and pericentric passage.  There are also examples of systems (plotted in yellow and orange) that appear to quench close 
to their first infall and pericentric passages.

To better understand these systems and examine their properties, we divide them into three categories: prompt quenchers that quench 
within 1 Gyr of first infall ($|t_{\rm{infall}}^{\rm{init}} - \tau_{90}| < 1$ Gyr), early quenchers that quench more than 1 Gyr before first infall 
($t_{\rm{infall}}^{\rm{init}} - \tau_{90} < -1$ Gyr), and late quenchers that quench more than 1 Gyr after first infall 
($t_{\rm{infall}}^{\rm{init}} - \tau_{90} > 1$ Gyr).  We also consider the population of unquenched systems.  

Figure \ref{fig:massesquenched} shows mass distributions for these four categories of satellites.  In general, unquenched systems occupy 
larger mass dark matter haloes than quenched systems and are more luminous.  It is also the case that the mean subhalo mass (either at 
infall or at the peak of the subhalo's history) for late quenchers is greater that that of prompt quenchers, and the mean mass of prompt 
quenchers is greater than that of the early quenchers.  Despite this, there is substantial overlap between the prompt quenchers and both 
the early and late quenchers in their subhalo mass distributions.  The subhalo masses of prompt quenchers encompass almost the entire 
range of subhalo masses for both early and late quenchers.  These trends indicate that while mass plays an important role in mediating the 
time of quenching, it is not the only factor driving quenching in these systems.

\subsection{Backsplash systems}
\label{sec:backsplash}
The previous section demonstrated that surviving satellites within \R\ can have complex interactions with their hosts over many Gyrs.  In 
this section, we expand our analysis to satellites beyond \R\ (out to 1 Mpc) and explore other metrics for quantifying `infall time.'

There are several categories of systems in the volume that extends out to 1 Mpc: systems that have never entered the host \R\ (35\% of all 
systems); systems that enter the host \R\ at high redshift, but exit again, and do not return by redshift zero (24\%); systems that enter the 
host \R\, exit again, but do return by redshift zero (22\%); and systems that enter \R\ (typically at later times) and remain within \R\ until 
redshift zero (19\%).  Another way of quantifying the exchange of systems across the host's \R\ that occurs over the course of the 
simulation is to say that of the systems beyond \R, but within 1 Mpc at $z=0$, 41\% have at some point crossed within \R\ prior to $z=0$, 
and of the systems within \R\ at $z=0$, 53\% have spent some amount of time after first infall outside \R.

\begin{figure}
\centering
\includegraphics[width=\columnwidth]{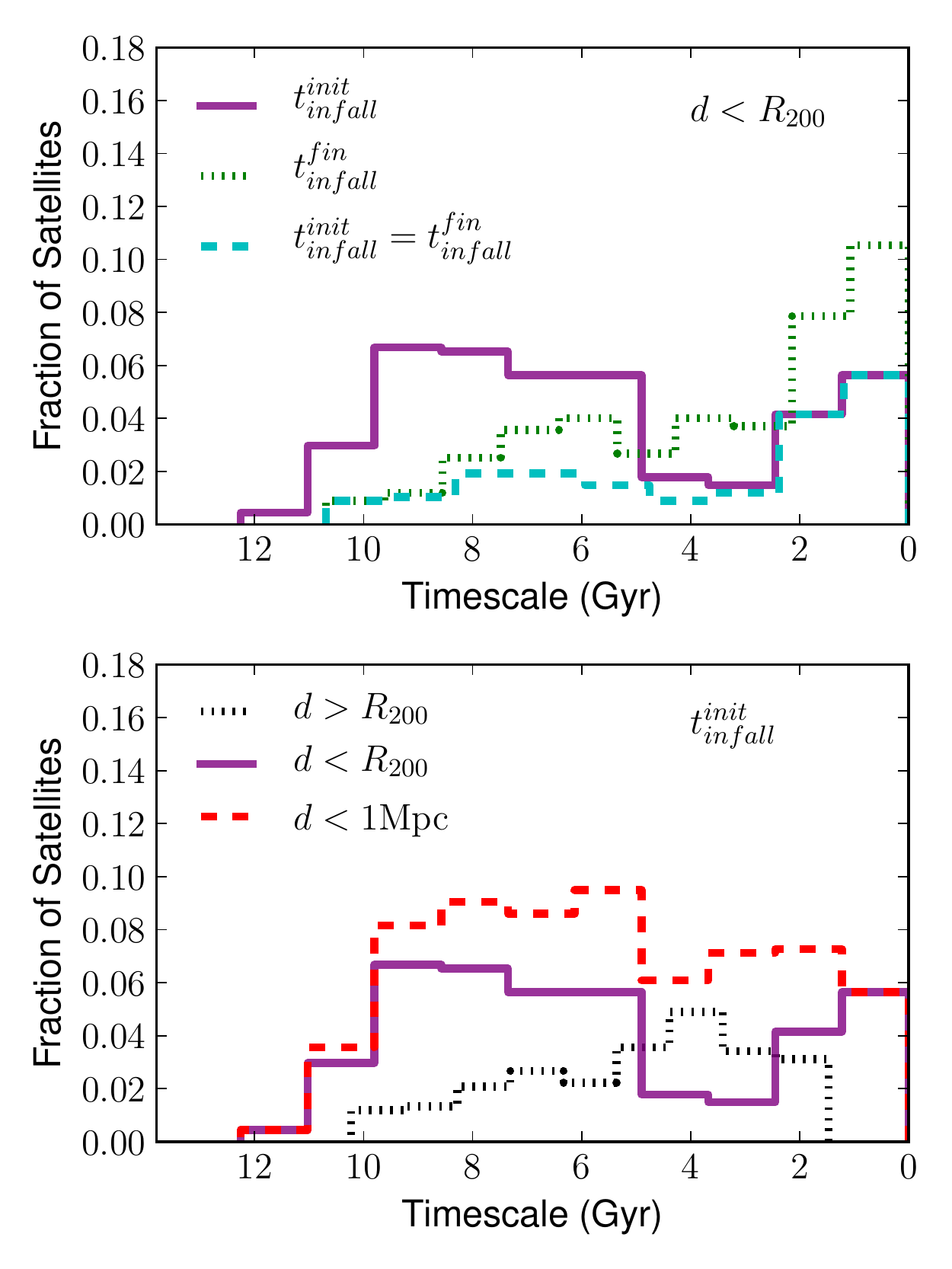}
\caption{Top: Distributions of infall lookback times for satellites that end the simulation within \R.  Two different definitions of infall time are explored: 
$t^{\rm{init}}_{\rm{infall}}$, the time when the satellite first crosses \R, and $t^{\rm{fin}}_{\rm{infall}}$, the final time the satellite crosses 
within \R.  Distributions of these timescales are shown, along with the distribution of satellites for which
$t^{\rm{init}}_{\rm{infall}}$ and $t^{\rm{fin}}_{\rm{infall}}$ are equal.  These systems cross within \R\ once and remain 
within \R\ until the end of the simulation.  The fraction of satellites plotted is the number of satellites in each bin 
divided by the total number of satellites within 1 Mpc (673).  Bottom: Distributions of $t^{\rm{init}}_{\rm{infall}}$ for systems in different 
distance samples.  Systems beyond \R\ that are included in this plot are backsplash satellites: systems that have been within \R\ of the 
main host sometime during their history, but are beyond \R\ at $z=0$.   }
\label{fig:infall_times}
\end{figure}

Figure \ref{fig:infall_times} compares the distributions of the initial infall time ($t^{\rm{init}}_{\rm{infall}}$) and the final infall time 
($t^{\rm{fin}}_{\rm{infall}}$), which is the time when satellites within \R\ last crossed \R.  This plot shows the bi-modal distribution of initial 
infall times, but now it is apparent that the late-infall peak is composed exclusively of systems that have only crossed within \R\ one time and 
remained within \R\ for the rest of the simulation.  For these systems, the initial and final infall times are equal 
($t^{\rm{init}}_{\rm{infall}} = t^{\rm{fin}}_{\rm{infall}}$).  

There are some early-infalling systems that also remain within \R\ for the entire simulation, but most of the early-infall peak is composed of 
returning `backsplash galaxies,' i.e. systems that had their first encounter with the host galaxy at high redshift, but have only recently 
returned to the host.  The quantity \R\ does grow with time as the universe expands and the host haloes increase in mass, but for many hosts, 
the growth in \R\ is within a factor of two over the past 8 Gyrs.

A result of this pattern is that the deficit of systems with initial infall times around 4 Gyrs ago can be accounted for by expanding this 
analysis to systems beyond \R.  Among systems beyond \R\ (but within 1 Mpc) that have had an interaction with the host halo, the most 
common initial infall time lies within the gap found among systems within \R.  This is shown in Figure \ref{fig:infall_times}.  The distribution of 
initial infall times among all systems within 1 Mpc does not exhibit the same degree of bi-modality.

\begin{figure}
\centering
\includegraphics[width=\columnwidth]{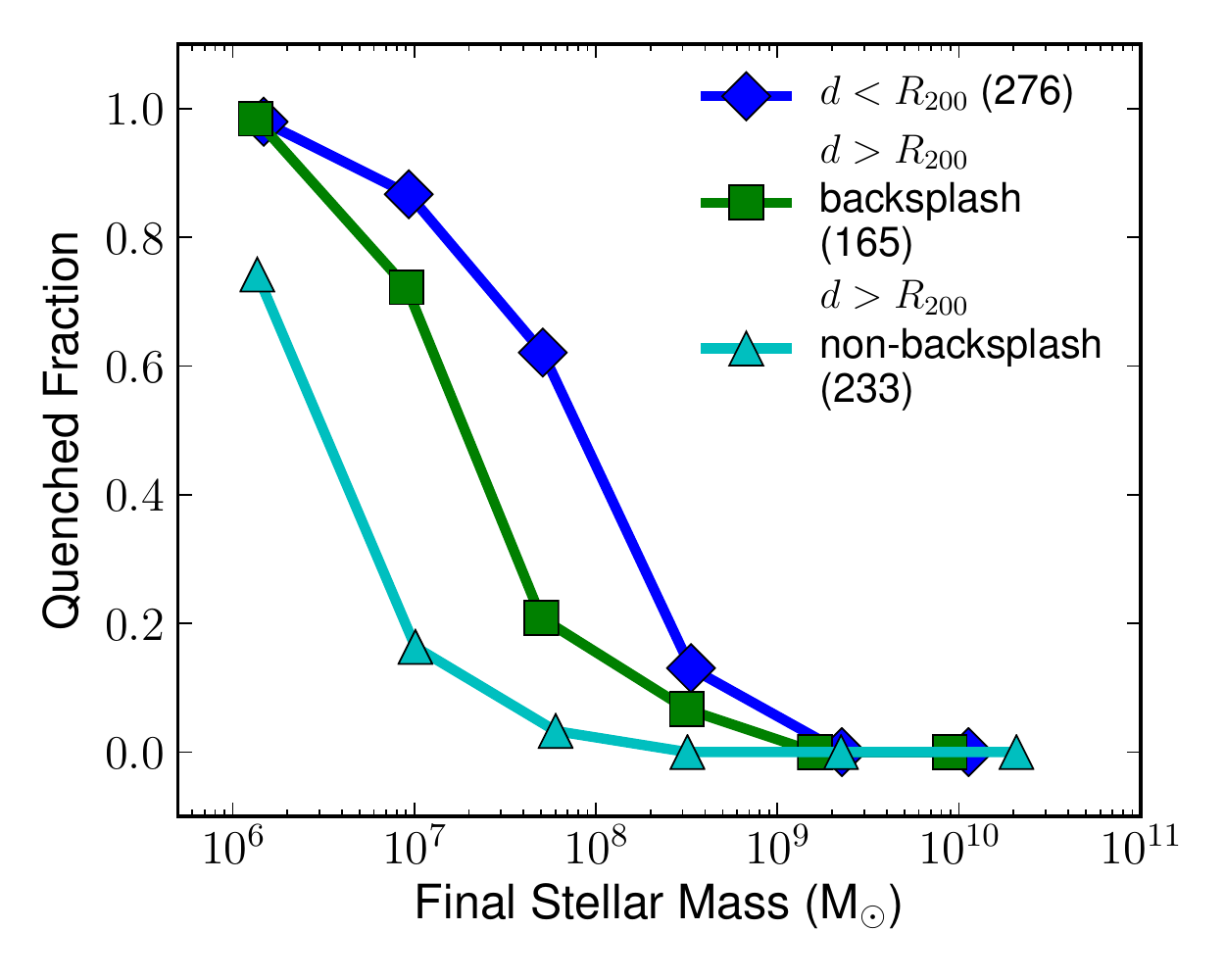}
\caption{The fraction of quenched subhaloes versus final stellar mass at redshift zero.  The trend is shown for three different samples of 
systems: subhaloes within \R\ at redshift zero (blue diamonds), subhaloes beyond \R\ at redshift zero, but which have been within the host
\R\ at some time in the past (green squares), and subhaloes beyond \R\ that have never been within \R\ (cyan triangles).  The number of 
systems in each sample is indicated.}
\label{fig:qfrac_infall}
\end{figure}

Figure \ref{fig:qfrac_infall} shows that backsplash systems beyond \R\ demonstrate quenching differences as compared to systems of the 
same stellar mass that have never interacted with the host.  The difference between systems in these two groups is most pronounced at a 
stellar mass of $10^7$ \Msun, where only 20\% of systems that have never interacted with the host are quenched versus a quenched 
fraction of 70\% for backsplash systems.  Indeed, backsplash systems beyond \R\ share a quenching distribution that is very similar to 
systems within \R.  

\section{Discussion}
\label{sec:discussion}

\subsection{Quenching in the Local Group and its environs}

How does star formation in the simulated satellite systems presented here compare to the satellite systems of the Local Group?  It should 
first be noted that the Auriga simulations model haloes similar to the MW in terms of mass and central galaxy properties, but the overall 
environment of the Auriga haloes (within 1 Mpc) is quite different from the real MW, which has the close neighbor M31 $\sim$ 700 kpc 
away.  There are likely circumgroup environment effects in addition to the circumgalactic environment.

Despite this difference, some aspects of Local Group satellite systems are reproduced.  Within 300 kpc of the MW, only two (the LMC and 
SMC) satellite galaxies are unquenched, out of a total of 10 known systems above our stellar mass cut.  This yields an overall quenched 
fraction of 80\%.  All systems with a stellar mass at or above $4.6 \times 10^7$ \Msun\ (the stellar mass of the SMC) are unquenched, also 
consistent with the Auriga satellites (see Figure \ref{fig:qfrac}).  Within M31, there is a similar picture: within 300 kpc, the overall quenched 
fraction is greater than 80\%.  These quenched fractions are broadly consistent with the degree of quenching found in the Auriga sample; 
within 300 kpc, 13 of the Auriga host haloes have quenched fractions above 80\% and 17 have quenched fractions below 80\%.  There may 
be undiscovered satellites within the MW and M31.  These systems are more likely to be quenched, given the low surface brightness of dSphs, 
and would increase the overall quenched fraction for each host.  The Auriga satellite systems are quite different from the satellite systems of the 
SAGA survey \citep{Geha2017} that finds uniformly star forming satellite systems in 8 MW analog galaxies down to a satellite luminosity of $M_r = -12.3$.

Our simulations produce a very steep quenching trend with stellar mass with a quenched fraction of $\sim$30\% at $M_{star} = 10^8$ \Msun\ and 
a quenched fraction $\sim$90\% at $M_{star} = 10^6$ \Msun.  Observational studies of satellite galaxies at these luminosities do not present a 
consistent picture of this trend; however, most find a less steep trend than simulated here.  \citet{Weisz2015} find a much shallower slope within 
the Local Group from an analysis of resolved stellar populations, with an especially high quenched fraction at $M_{star} = 10^8$ \Msun\ of over 
50\%.  \citet{SlaterBell2014} also find a shallower quenching trend with mass with Local Group data from the NASA-Sloan Atlas, but do find the shift in 
quenched fraction to occur at a stellar mass of $10^7$\Msun, consistent with our results.  The data of \citet{Karachentsev2013} of galaxies within 
the local volume ($\le 11$ Mpc) give a much steeper slope for satellite galaxies based on galaxy types, with a lower quenched fraction at 
$10^8$ \Msun, but the quenched fraction at $10^6$ \Msun\ is less than 70\%, lower than the 90\% quenched fraction in Auriga \citep{Weisz2015}.  
It should be noted that both samples lack completeness and apply to environments not directly analogous to the environment simulated here, 
especially in the case of \citet{Karachentsev2013}, whose sample includes satellites around a more heterogeneous group of hosts.

Our simulations also produce a quenching trend with distance (Figure \ref{fig:qfrac}) and the Local Group does demonstrate a correlation between 
environment and galaxy type, with dSphs being more clustered around the MW and M31 and dIrrs being more common at larger distances (with the 
exception of the Magellanic Clouds).  \citet{Geha2012} found that quenched field galaxies between $10^7$ \Msun\ and $10^9$ \Msun\ are 
extremely rare, consistent with our findings; however, our most isolated systems are not as isolated as their field sample.

\begin{figure}
\centering
\includegraphics[width=\columnwidth]{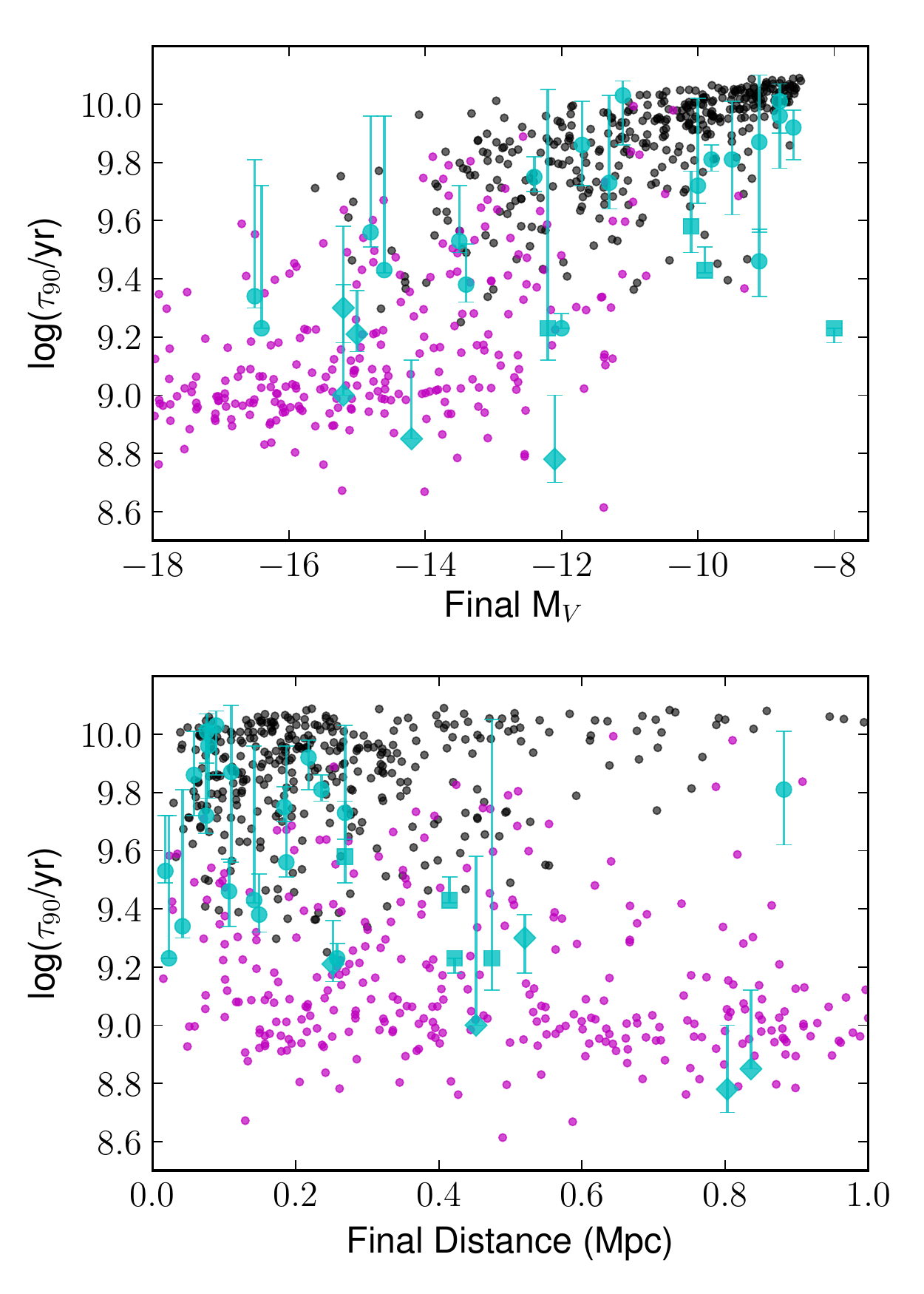} 
\caption{Timescale $\tau_{90}$ plotted versus V band absolute magnitude (top panel) and final distance from host (bottom 
panel).  $\tau_{90}$ is the lookback time when a system has formed 90\% of its stellar mass.  All satellites in our sample (systems that 
have a final halo mass above $10^8$ \Msun, a final stellar mass above $5 \times 10^5$ \Msun, and a final distance less than 1 Mpc) are 
plotted.  Quenched systems are plotted in black and starforming systems are plotted in magenta.  Observational measurements of $
\tau_{90}$ for Local Group systems from \citet{Weisz2015} are plotted in cyan.  We include systems with $M_V \le -8$ and a distance to 
the nearest large galaxy (either the MW or M31) less than 1 Mpc.  Early type systems (dSphs and dwarf ellipticals) are plotted with circles; 
dwarf irregulars are plotted with diamonds; and transitional dwarfs (systems with evidence of recent star formation in their 
color-magnitude diagrams, but no H$\alpha$ emission) are plotted with squares.  }
\label{fig:obs}
\end{figure}

In terms of quenching times, our sample is consistent with observational measurements of $\tau_{90}$ from resolved stellar population 
studies \citep{Weisz2015}.  Figure \ref{fig:obs} shows this comparison and, like observed systems, there is a dearth of high-luminosity, 
early-$\tau_{90}$ systems and low-luminosity, late-$\tau_{90}$ systems.  There are a few notable inconsistencies.  While observed dSphs 
and dEs tend to lie with simulated quenched systems and observed dTrans and dIrr systems tend to lie with simulated 
unquenched systems, the two prominent dEs of M31, M32 and NGC 205, have values of $\tau_{90}$ later than most quenched systems in 
our sample and that are more consistent with our unquenched systems. Also, while low-luminosity quenched systems in our sample have 
values of $\tau_{90}$ consistent with dSphs of similar luminosity, they tend to quench somewhat earlier than the typical observed system.  
In addition, although our sample does not extend down to the luminosity of Leo T, Leo T has an unusually late $\tau_{90}$ value for its 
luminosity with respect to both other observed systems and the trend of our simulated sample.

These results combined with our characterization of satellite infall times, confirms previous modeling of dark matter only simulations that found that short 
quenching timescales relative to satellite infall times (less than 2 Gyr) are required to reproduce the quenched state of Local Group satellite 
galaxies \citep{SlaterBell2014,Wetzel2015,Fillingham2015}. 

We found in Auriga a number of `backsplash' satellite galaxies and demonstrated that their quenching properties are distinct from field systems of similar mass 
that have never had an encounter with the central host (Figure \ref{fig:qfrac_infall}).  Previous work done with cosmological nbody simulations have also found 
populations of backsplash dwarf galaxies in a Local Group context and demonstrated that they have distinct orbital dynamics from other field 
populations \citep{Teyssier} and can explain the properties of smaller galaxies surrounding groups and clusters \citep{Wetzel2014}.

\subsection{The physics of ram pressure stripping and quenching}

Comparisons to the Local Group suggest that despite our model's success at capturing general trends, subhaloes in our sample are perhaps 
under-quenched at high masses and over-quenched at low masses.  For low-mass systems, this over-quenching appears to happen at 
early times.

A likely explanation of this trend is the stellar feedback and ISM model employed in the Auriga simulations.  The Auriga simulations use the subgrid ISM model 
of \citet{SpringelHernquist2003} in which dense molecular gas (and low-temperature cooling) is not directly simulated; rather, molecular gas is assumed to be 
below the resolution of gas cells and its impact on the galaxy is treated in a `subgrid' fashion: it is assumed to be in pressure equilibrium of the hot phase of 
the ISM and this pressure is imposed for gas above the star-formation density threshold.  However, by not directly simulating molecular clouds, we are also 
not directly simulating the effects of ram pressure on the molecular phase of the ISM.
Idealized simulations of ram pressure stripping at high resolution that model a multiphase medium find that clumps with the density of molecular clouds are 
able to survive certain low-ram pressure circumstances and that lower density gas is stripped more easily \citep{TonnesenBryan2009}.

Several studies have explored the impact of star formation and internal feedback on ram pressure stripping and have found that it is a minor effect both on the 
cluster and Milky Way scales \citep{TonnesenBryan2012,Emerick}.  However, the feedback model in Auriga is also responsible for the properties and 
evolution of the CGM and likely impacts the effectiveness of ram pressure stripping by contributing to the density of the CGM.  Future studies will explore the 
CGM properties of the Auriga simulations.

Auriga's ISM and feedback model also appears to regulate a gas-star formation equilibrium for isolated subhaloes with a constant central gas fraction and a 
continuous SFR that scales with the halo mass (see Figure \ref{fig:halo_evo}).  Modeling stellar feedback effects in a realistic ISM may be necessary to 
capture the stochastic aspects of starbursts that may lead to quenching within higher-mass subhaloes, especially systems that are being `starved', i.e. 
systems that have their gas supply cut off due to their proximity to the larger host and are left to consume their existing gas supply.  Studies of the dynamics of 
Local Group satellites indeed suggest that there is a shift in the quenching mechanism of satellites above $10^8$ \Msun and that starvation, where the 
quenching time becomes the cold gas depletion time, is consistent with observations\citep{Fillingham2015, Wheeler2014}.  Auriga's ability to capture this 
mode of quenching is therefore dependent on the ISM and feedback model, which is not tuned to reproduce satellite quenching itself, but rather the cosmic 
star formation history and abundance matching constraints for central galaxies, and lacks a bursty mode of star formation.

\subsection{Satellite quenching in different environments}

Previous simulations have explored quenching and gas depletion in satellite galaxies in more massive host haloes, including galaxy clusters and groups.  
These studies typically do not have the baryon resolution to capture the low-mass satellites that we examine, but are able to examine quenching processes in 
more massive host halos and out to larger host distances.  \citet{Bahe2013} found environmental quenching around hosts above $10^{13}$ \Msun\ in mass 
out to 5\R.  The degree of quenching was dependent on the host mass, satellite mass, and the satellite distance.  At their lowest host mass, $10^{13}$\Msun\ 
in total mass, and lowest satellite mass, $10^9$ \Msun\ in stars, they found quenching began within 3\R\ and reached over 80\% within \R.  Our simulations 
found no quenching in satellites of this luminosity, perhaps indicating a sharp break in quenching behavior in host haloes between 
$10^{12}$ and $10^{13}$ \Msun.  In a follow-up study, \citet{BaheMcCarthy2015} found quenching of their lowest luminosity satellites ($< 10^{10}$ \Msun in 
stars) is very rapid.  We also find rapid quenching for many systems, but it does not appear to be strictly a function of mass in our simulations.

At the scale of the MW, idealized simulations with a constant background CGM density have suggested that ram pressure alone may not be an effective 
quenching mechanism \citep{Emerick}, however, we note that the CGM density in our models changes by more than an order of magnitude between 
$\sim 10$ and $\sim 100$ kpc, and is clumpy and non-axis symmetric.  \citet{Fillingham2016} found that ram pressure in a clumpy CGM can be an 
effective quenching mechanism at the MW scale capable of quenching $\sim90\%$ of infalling satellites based on the properties of local field dwarfs.  
Future simulations with higher resolution in the CGM will allow us to test the effect of CGM clumpiness in greater detail.

\section{Conclusions}
\label{sec:conclusions}

In this paper, we have presented a picture of field dwarf and satellite galaxy quenching as modeled in the Auriga cosmological zoom-in 
simulations of Milky Way type galaxies in $\sim 10^{12}$ \Msun\ haloes.  These simulations reproduce general properties of the satellite systems 
of the MW and M31 in terms of number and quenched fraction.  We find that ram pressure stripping is a dominant quenching mechanism 
and that approximately 50\% of satellite systems (systems within \R) quench within 1 Gyr of first infall with their host.

A summary of several key findings:
\begin{itemize}
\item The Auriga simulations have an average of 11.9 satellites with stellar masses above $5 \times 10^5$ within 300 kpc.  Subhaloes with 
this luminosity or greater exhibit a strong mass-dependent quenching pattern, with 90\% of systems at $M_{\rm{star}} \sim10^6$ \Msun\ 
being quenched regardless of distance.
\item Systems with $M_{\rm{star}} < 10^8$ \Msun\ demonstrate distance-dependent quenching.  The distance where increased quenching 
takes 
place shifts to smaller distances for lower masses.  For the most isolated systems in our sample (beyond 600 kpc), a sharp rise in the 
quenched fraction occurs below $M_{\rm{star}} \sim 10^7$ \Msun.
\item Ram pressure stripping is a ubiquitous phenomenon in lower-mass systems ($M_{star} < 10^7$ \Msun) and can produce total gas loss 
in these systems.  
\item Other quenching mechanisms we consider are cosmic reionization, stellar feedback, environmental effects, and tidal 
stripping.  We find that by selecting systems with a minimum stellar mass of $5 \times 10^5$ \Msun, our sample does not include 
reionization fossils.  Stellar feedback in our model regulates the gas content within dwarf haloes, maintaining a central gas fraction of 
$\sim$~10\%, but lacks the bursty nature to be a dominant quenching mechanism.  Larger-scale environmental effects (e.g. interactions 
with the cosmic web) appear to play an important role for some systems.  Tidal effects dramatically impact the dark matter content of satellite 
systems, but due to the concentration of stellar populations, tidal destruction of the stellar component occurs much later than the disruption 
of the dark matter.
\item Of systems within \R\ at $z=0$, 51\% of quenched systems (37\% of all systems) quench within 1 Gyr of first crossing their hosts' \R\ 
radius.  Systems that quench more than 1 Gyr after first infall tend to be more luminous and systems that quench more than 1 Gyr before 
first infall tend to be less luminous.
\item Systems within \R\ at $z=0$ exhibit a bi-modality in their infall times, indicative of a dual population within \R\ of newly in-falling 
systems and returning backsplash systems that have had a previous interaction with the host.  
53\% of systems within \R\ have spent some amount of time outside of \R\ after their first infall.
\item There is a significant population of backsplash systems that fell in earlier and are currently beyond \R.  The backsplash population shows quenching properties much like systems with \R.  
41\% of systems beyond \R\ but within 1 Mpc at $z=0$ are backsplash systems.
\item The simulated satellites and dwarf galaxies discussed here compare favorably to the systems of the Local Group.  One discrepancy is 
that our estimates for H~I gas mass are a factor of a few too high, but the quenching times of our simulated systems broadly agree with 
observed star formation histories.
\end{itemize}

This study focused on surviving dwarf galaxies in the vicinity of a MW type galaxy.  These systems are only a portion of the dwarf galaxies 
that have interacted with the host galaxies over time, and a study of the tidally disrupted systems that do not survive will address important 
questions such as how the satellites that built the stellar halo at high redshift compare to present day systems or how did dwarf galaxies 
contribute to reionization of the local volume.

This study also only touched on dwarf systems with sufficiently resolved stellar components in order to quantify quenching timescales.  
However, a majority of systems are dark and under-luminous, as shown in Figure \ref{fig:cum_distr}.  Quantifying this 
population will provide insights into mechanisms for galaxy suppression, not just quenching, and likely include reionization and perhaps 
environmental effects.

Observational work in this area is in an exciting time.  Our ability to probe H I in more diverse environments with the EVLA and SKA pathfinders will yield a 
more complete picture of gas in different environments beyond the local universe \citep{Fernandez2016}.  Just recently a ram-pressure stripping candidate 
around a local MW analog was discovered with the Westerbork Synthesis Radio Telescope \citep{Hsyu2017}.

This study has demonstrated the power that a large sample of cosmological models with full hydrodynamics can provide in quantifying 
trends in subhalo properties.  Each Auriga halo has a unique history, and while the Auriga haloes all have very similar halo masses, they 
can and do have quite a diversity in their substructure properties.  In order to understand the underlying physical mechanisms driving 
evolution in satellite systems as a whole, it is necessary to study them with sufficient statistical power.

\section*{Acknowledgements}
The authors would like to thank the anymous referee whose comments helped improve this paper.  
This work has been supported by the European Research Council under ERC-StG grant EXAGAL-308037, and by the Klaus Tschira Foundation.  
CMS would like to acknowledge helpful conversations with Antonela Monachesi, Greg Bryan, Kathryn Johnston, Mary Putman, 
Jacqueline van Gorkom, and Stephanie Tonnesen and comments from Adrian Jenkins.  RG and VS acknowledge support by the DFG Research Centre 
SFB-881 `The Milky Way System' through project A1. 
Part of the simulations of this paper used the SuperMUC system at the Leibniz Computing Centre, Garching, under the 
project PR85JE of the Gauss Centre for Supercomputing.
This work was supported by the Science and Technology Facilities Council (grant numbers ST/F001166/1, ST/I00162X/1) and European Research Council 
(grant number GA 267291 ``Cosmiway''). It used the DiRAC Data Centric system at Durham University, operated by the Institute for Computational Cosmology 
on behalf of the STFC DiRAC HPC Facility (www.dirac.ac.uk). This equipment was funded by BIS National E-infrastructure capital grant ST/K00042X/1, STFC 
capital grant ST/H008519/1, and STFC DiRAC Operations grant ST/K003267/1 and Durham University. DiRAC is part of the National E-Infrastructure.
    
\bibliographystyle{mnras}
\bibliography{paper}

\appendix
\section{Resolution and convergence}

In this appendix, we explore the impact of numerical resolution on the satellite system properties.  We do this by comparing the 30 Auriga zoom simulations 
discussed in the main text to six resimulations of Auriga halos with a factor of 8 better mass resolution as described in \citet{Grand2016}.  We especially focus 
on lower mass satellite systems since they have fewer resolution elements.

\begin{table*}
\caption{Subhalo population properties by host halo for the high-resolution Level 3 simulations.}
\label{tab2}
\begin{tabular}{lccccccc}
\hline
Simulation & Host $M_{200}$     & Host $R_{200}$ & $N_{\rm{sub}}$  & max($v_{\rm{max}}$) & $f_{\rm{quenched}}$    
  & $f_{\rm{HIpoor}}$ & min($M_{H I} > 0$)\\
 & ($10^{12}$ \Msun) & (kpc)                   &   (< 300 kpc/1 Mpc)                & (km s$^{-1}$)       &  (< 300 kpc/1 Mpc)   
 &(< 300 kpc/1 Mpc)  & ($10^{6}$ \Msun)  \\
  \hline

Au6 & 1.01 & 212 & 8/13 & 93 & 0.5/0.38 & 0.5/0.31 & 2.14 \\ 

Au16 & 1.5 & 242 & 15/29 & 117 & 0.67/0.48 & 0.67/0.45 & 0.74\\ 

Au21 & 1.42 & 237 & 17/35 & 108 & 0.65/0.46 & 0.65/0.43 &  0.82\\ 

Au23 & 1.5 & 242 & 16/22 & 51 & 0.75/0.64 & 0.81/0.64 & 8.13 \\ 

Au24 & 1.47 & 240 & 14/37 & 120 & 0.64/0.41 & 0.57/0.32 &0.85 \\ 

Au27 & 1.7 & 251 & 13/18 & 113 & 0.69/0.61 & 0.69/0.61 & 1.55 \\ 
 \hline     

 \end{tabular}

\raggedright  Note: The quantities presented in each column are the same as Table \ref{tab} except for the final column, which is the lowest positive H I mass 
found in satellites of the host.

\end{table*}

Table \ref{tab2} lists the satellite properties of the six zoom simulations from our halo suite that have a target cell mass of $6 \times 10^3$ \Msun\ and 
minimum softening length of 184 pc at $z=0$ for high resolution dark matter particles and star particles.  We call these simulations `Level 3' resolution 
simulations and the standard resolution simulations listed in Table \ref{tab} `Level 4' resolution simulations. 

The stellar mass threshold we adopt in the main text is 10 times the typical mass of a star particle at Level 4 resolution.  With 8 times better mass resolution, 
systems at this stellar mass limit in the Level 3 runs should have approximately 80 star particles.

\begin{figure*}
\centering
\includegraphics[width=\textwidth]{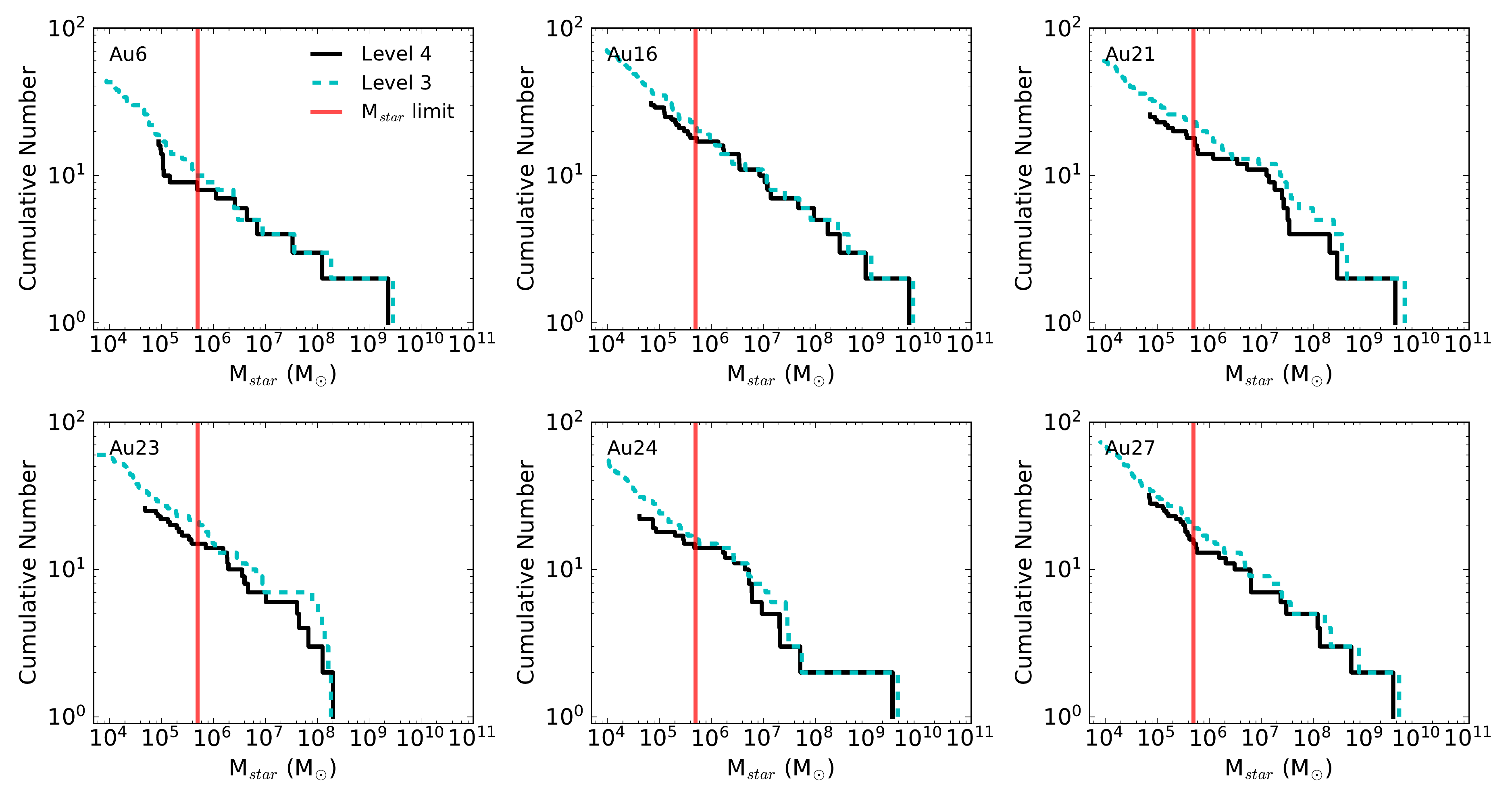}
\caption{Cumulative distributions of the final satellite stellar masses for Auriga host halos simulated at both the high Level 3 mass resolution (dashed cyan) 
and the low Level 4 mass resolution (solid black).  The mass resolution in the two sets of simulations differ by a factor of 8.  The stellar mass limit 
($5 \times 10^5$ \Msun) adopted for the analysis in the main text of the paper is shown with a vertical red line.  }
\label{fig:cdist_convergence}
\end{figure*}

Figure \ref{fig:cdist_convergence} shows how the luminosity distributions compare between haloes simulated with the same initial conditions, but with different 
mass resolution.  There are small differences in the stellar masses of some satellites, but the overall luminosity distributions follow each other between 
resolution levels.  The final numbers of satellites are similar down to the stellar mass threshold we adopt, with only a slight excess of satellites in the Level 3 
simulations.  

\begin{figure*}
\centering
\includegraphics[width=\textwidth]{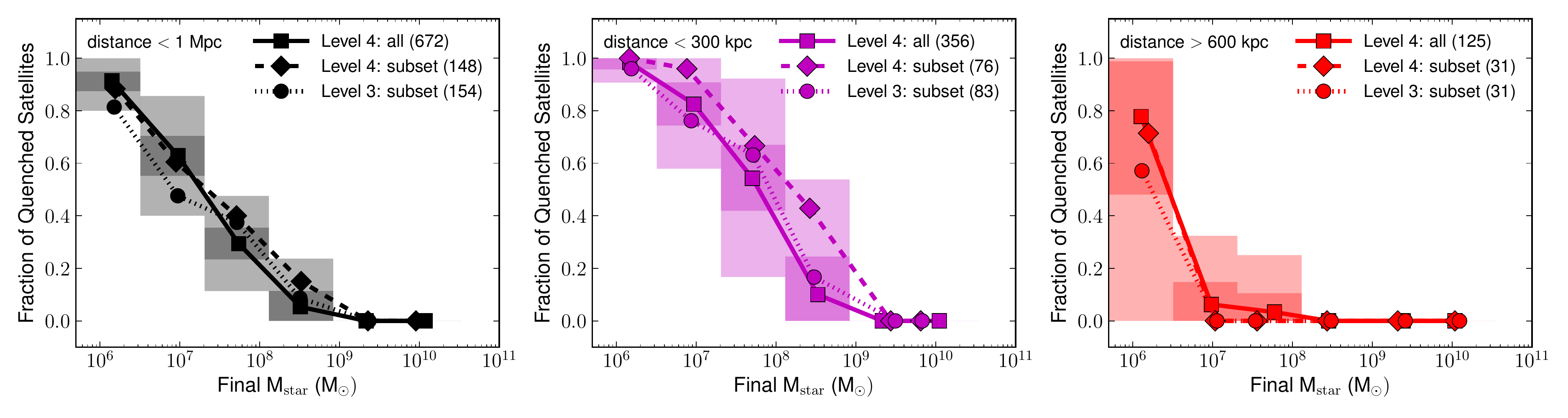}
\caption{Comparison of the quenched fractions of dwarf systems as a function of final stellar mass between the Level 4 simulations and the higher-resolution 
Level 3 simulations.  We separate dwarf systems into three distance bins in the three panels shown: all systems within 1 Mpc (black, left panel), systems 
within 300 kpc (magenta, center panel), and systems beyond 600 kpc, but within 1 Mpc (red, right panel).  The quenched fraction of systems in all thirty Level 
4 haloes (solid lines with squares) are shown in the same way as in Figure \ref{fig:qfrac} in each distance bin.  Trends for all systems in the six Level 3 halos 
are shown (dotted lines with circles) and trends for all systems in the six Level 4 haloes that match the Level 3 haloes (dashed lines with squares).  Shaded 
regions show the range of quenched fractions produced by $10^4$ randomly selected sets of six distinct Level 4 haloes.  The light shaded regions show the 
three-sigma range in the quenched fraction around the mean and the dark shaded regions show the one-sigma range.  Higher-mass bins do not show a range 
of values because, above a certain stellar mass, all systems are star-forming (as described in Section \ref{sec:sfproperties}).     }
\label{fig:qfrac_convergence}
\end{figure*}

Figure \ref{fig:qfrac_convergence} compares the quenched fractions of dwarf galaxies between the Level 3 and Level 4 resolution simulations.  The 
characteristic up-turn in the fraction of quenched systems with stellar masses below $10^7$ \Msun\ is apparent in the Level 3 simulations to a similar degree 
as the Level 4 simulations.  It should be noted however, that at large distances especially, the smaller number of dwarf systems in the Level 3 simulations 
introduces a larger scatter in fractional results.  From this analysis, we conclude that the evolution of low mass systems is numerically converged with our 
model.

Table \ref{tab2} lists the lowest non-zero H~I masses found in satellites of each host halo within 1 Mpc.  These values are computed with the same methods 
and limits as the results in the main section of the text.  These values are similar to the lowest values found in the Level 4 simulations.  The additional of 
resolution of the Level 3 runs does not allow us to capture satellites with lower H I masses, mainly because this calculation relies on assumptions of our 
subgrid model and is therefore insensitive to resolution.

\label{lastpage}

\end{document}